\documentclass[3p,times,review]{elsarticle} 

\usepackage[dvipsnames,table]{xcolor}

\usepackage{amsmath}

\usepackage{siunitx}
\usepackage{ulem}
\usepackage[ruled,linesnumbered]{algorithm2e}

\usepackage{booktabs}
\usepackage{subcaption}

\usepackage{graphicx}
\usepackage{subcaption}
\usepackage{adjustbox}
\usepackage{multirow}
\usepackage{array,makecell}
\usepackage{listings}
\usepackage{amsmath}
\usepackage[inline]{enumitem}

\usepackage{hyperref}

\usepackage{amssymb}

\journal{Additive Manufacturing}

\begin{document}

\begin{frontmatter}



\title{One-Shot Camera-Based Extrusion Optimization for  \\High Speed Fused Filament Fabrication}


\author[inst1,inst4]{Yufan Lin}
\ead{yuflin@student.ethz.ch}
\author[inst1,inst2]{Xavier Guidetti\corref{cor1}}
\ead{xaguidetti@control.ee.ethz.ch}
\author[inst3]{Yannick Nagel}
\ead{yannick.nagel@nematx.com}
\author[inst1,inst2]{Efe C. Balta}
\ead{efe.balta@inspire.ch}
\author[inst1]{John Lygeros}
\ead{lygeros@control.ee.ethz.ch}

\affiliation[inst1]{organization={Automatic Control Laboratory, ETH Zurich},
            addressline={Physikstrasse 3}, 
            postcode={8092}, 
            city={Zurich},
            country={Switzerland}}
            
\affiliation[inst2]{organization={Inspire AG},
            addressline={Technoparkstrasse 1}, 
            postcode={8005}, 
            city={Zurich},
            country={Switzerland}}
            
\affiliation[inst3]{organization={NematX AG},
            addressline={Vladimir-Prelog-Weg 5}, 
            postcode={8093}, 
            city={Zurich},
            country={Switzerland}}
            
\affiliation[inst4]{organization={Department of Mechanical and Process Engineering, ETH Zurich},
            addressline={Leonhardstrasse 21}, 
            postcode={8092}, 
            city={Zurich},
            country={Switzerland}}
            
\cortext[cor1]{Corresponding author}

\begin{abstract}
Off-the-shelf Fused Filament Fabrication 3D printers are widely accessible and convenient, yet they exhibit quality loss at high speeds due to dynamic mis-synchronization between printhead motion and material extrusion systems, notably corner over‑extrusion. Existing methods require specialized hardware, extensive calibration, or firmware modifications that are inaccessible to most users. This work presents a practical, end-to-end optimization framework that enhances high-speed printing using only standard 3D printers and a phone camera, without requiring additional complex setup. 
The method employs a one-shot calibration approach where two simple printed patterns, captured by a phone camera, enable identification of extrusion dynamics and cornering behavior. Indentified systems enable a model-based constrained optimal control strategy that generates optimized G-code, synchronizing motion and extrusion.
Experiments show reduced width tracking error, mitigated corner defects, and lower surface roughness, achieving surface quality at \SI{3600}{\milli\meter\per\minute} comparable to conventional printing at \SI{1600}{\milli\meter\per\minute}, effectively doubling production speed while maintaining print quality. This accessible, hardware-minimal approach enables a wide range of FFF users to achieve high-quality, high-speed additive manufacturing.

\end{abstract}




\end{frontmatter}

\section{Introduction} \label{sec:intro}
\subsection{Background} \label{subsec:background} 
\begin{figure}[ht]
     \centering
     \includegraphics[scale=1]{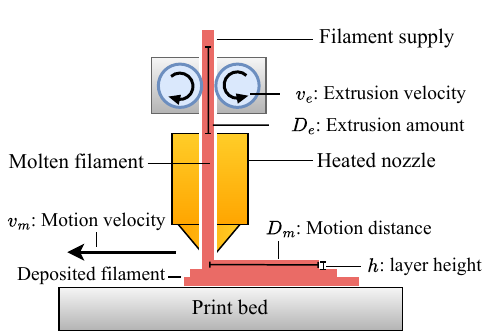}\\
     \caption{The process of 3D printing}
     \label{fig:3DPrintingProcess}
\end{figure}

Material extrusion Additive Manufacturing (AM), also known as Fused Filament Fabrication (FFF), has become one of the most popular 3D printing techniques for thermoplastics. This layer-wise manufacturing process offers numerous advantages, such as design flexibility, rapid prototyping and easy accessibility of hardware and software. This has attracted a large number of users, both hobbyists and professionals, making FFF widely used in a very broad range of applications. Despite its apparent simplicity, FFF is a complex nonlinear process composed of several interconnected sub-processes~\cite{iso2021additive}. As shown in Fig.~\ref{fig:3DPrintingProcess}, the printing process consists of the heating and melting of the materials, the molten materials extrusion and deposition, and the traveling motion of the printhead. This complexity leads to challenges in parameters tuning and control design for the printing process. Currently, most desktop 3D printers utilize an open-loop control strategy, which produces acceptable results under ideal operation conditions. However, 3D printers have limited performance when the real machine behavior deviates from the expected one, for example when printing at very high speed. A number of process optimization methods~\cite{delfs2016optimized, dey2019systematic, kantaros2021employing, guidetti2023data} have been developed in academia or industry, utilizing high-performance professional equipment and sensors. However, these hardware intensive methods are not accessible for most home users of 3D printing. Developing an easily applicable process optimization method based on simple hardware would enable a vast number of FFF users to improve the performance of their printer under demanding manufacturing conditions.

This work aims to control material extrusion and deposition through a practical approach for common printer users. During printing, material is extruded and deposited along the printhead's commanded G-code path, shaping the bead cross-section and further impacting print quality~\cite{n2014review,turner2015review}. Numerical models~\cite{comminal2018numerical, serdeczny2018experimental, gosset2020experimental, balta2022numerical} have been developed and experimentally validated showing that the ratio of the material extrusion velocity to the nozzle motion velocity significantly influences the bead shape, determining the bead width when printing at a fixed layer height (i.e. the gap between nozzle and substrate). These studies reveal the steady-state effects of the velocity ratio, but transient dynamics—influenced by machine motion and extrusion systems behavior—also impact print quality and must be considered in control design.

\begin{figure}[ht]
  \centering
  \begin{subfigure}[b]{0.25\textwidth}
    \includegraphics[width=\textwidth]{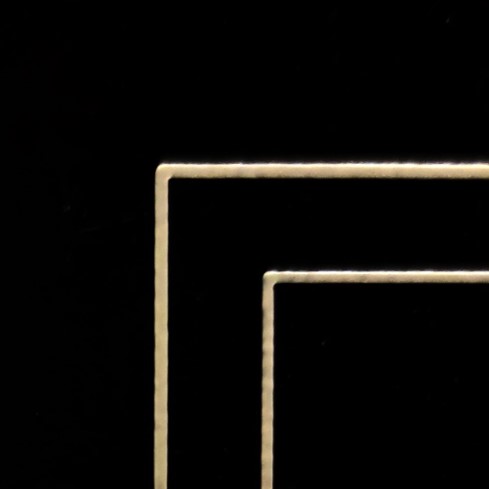}
    \caption{Corner Printed at \SI{900}{\milli\meter\per\minute}}
    \label{fig:Corner900}
  \end{subfigure}
  \hspace{1cm}
  \begin{subfigure}[b]{0.25\textwidth}
    \includegraphics[width=\textwidth]{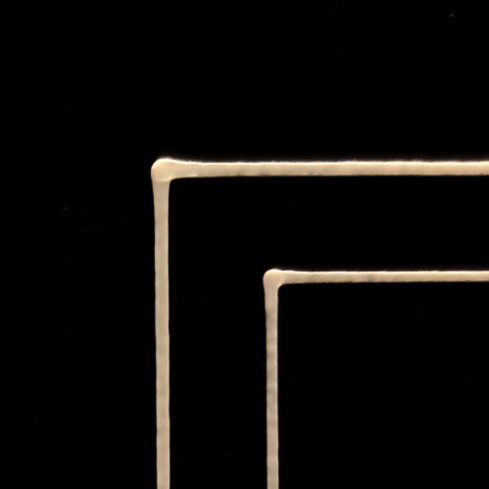}
    \caption{Corner Printed at \SI{3600}{\milli\meter\per\minute}}
    \label{fig:Corner3600}
  \end{subfigure}
  \caption{Comparison of corner printing at different speed}
  \label{fig:CornerCompare}
\end{figure}

The motion and extrusion systems exhibit distinct transient responses characterized by different time constants~\cite{bellini2004liquefier, n2014review}: the motion system responds rapidly, whereas the extrusion system responds slowly due to the extrusion dynamics of the viscous filament. This mismatch leads to poor synchronization in two systems, especially during high acceleration or deceleration, causing printing defects~\cite{barton2011cross}. A common defect is over-extrusion in corners when printing at high speeds, where the extrusion system fails to match abrupt velocity changes in the motion system. As Fig~\ref{fig:CornerCompare} shows, high speed corners, such as Fig~\ref{fig:Corner3600}, suffer from over-extrusion compared to those in standard print speeds shown in Fig.~\ref{fig:Corner900}. These defects not only produce irregular blobs shapes on corners, but also disrupt the deposition of the subsequent layer, deteriorating overall print quality. The defects severity increases with printing speed, which limits high-speed printing performance of current 3D printers and leading to slower processing times to achieve desirable print quality.

To coordinate the motion system and extrusion system, in the face of unknown nonlinear transient dynamics of the extrusion system, control strategies have been developed, ranging from trajectory optimization to online and feedforward control. Trajectory optimization focus on ensuring smooth printhead motion with continuous velocity~\cite{ertay2018synchronized, comminal2019motion}, aiming to reduce mis-synchronization by rapid acceleration, and extrusion is regulated proportionally. However, this smoothing may compromise geometric accuracy and reduce printing speed. Additionally, the slow response of the extrusion system hinders precise proportional control, highlighting the need for further control and optimization based on system analysis.

To advance the understanding and control of the extrusion process in additive manufacturing, researchers have increasingly employed in-situ measurement techniques using a variety of sensors, including cameras~\cite{cheng2008vision, hoelzle2008iterative, barjuei2022real, rojas2024model, farjam2024framework,sadaf2025spatiotemporal}, laser sensors~\cite{lu2014model, balta2021layer}, pressure sensors~\cite{adib2021hybrid}, and force/torque sensors~\cite{guidetti2024force, guidetti2024data}. Among these, cameras are particularly attractive due to their accessibility and versatility, and have been extensively used for in-situ monitoring and feedback control to detect and correct bead shape defects~\cite{jin2019autonomous, wang2020cnn, petsiuk2022towards, brion2022generalisable, barjuei2022real, rojas2024model}. Beyond in-situ measurements feedback, layer-to-layer (L2L) control strategies have also been explored. For example, L2L spatial height control has been implemented using model predictive control~\cite{lu2014model, guo2016predictive, inyang2020layer} and iterative learning control (ILC)~\cite{aksoy2020control, balta2021layer, afkhami2023robust, balta2024iterative, farjam2024framework, bahrami2025optimal}. Similarly, by analyzing bead width and defects in the previous layer with cameras, L2L fuzzy process control~\cite{cheng2008vision} and ILC~\cite{hoelzle2008iterative} have been used to optimize the flow rate for subsequent layers.  While these online feedback approaches can improve print quality, their reliance on additional hardware increases cost and complexity, limiting their accessibility for typical users.

To address the impracticality of online feedback acquisition, feedforward control methods based on parameter tuning or system identification have been developed. A widely adopted method in the 3D printing community is pressure advance (or linear advance), which adjusts the extrusion command based on an experimentally determined gain to pre-compensate for the lag in the extrusion system~\cite{PressureAdvance, LinearAdvance, tronvoll2019investigating}. While these techniques are easy to implement and effective in practice, they rely on time-consuming trial-and-error experimental tuning and lack a deep understanding of the coupled dynamics between extrusion and motion, leading to limited accuracy and generalizability~\cite{tronvoll2019investigating}.

Model-based feedforward control, grounded in system identification,has been widely adopted for printing process control. Approximating the the extrusion system as a first-order model, model-inverse feedforward control has shown effectiveness in both simulations~\cite{han2007coordination} and practical applications~\cite{chesser2019extrusion}. Iterative learning feedforward control~\cite{wu2021accurate} further improves width tracking by leveraging nonlinear extrusion models calibrated through multiple experiments with fixed camera setups. Additionally, phenomenological models~\cite{wu2023modeling} have been developed to enhance extrusion and retraction control. More recently, model-based optimal control strategies have emerged, utilizing either state-space models identified from experimental data~\cite{fravolini2025data} or physics-based formulations~\cite{bahrami2025optimal} to regulate line width. While these methods achieve state-of-the-art performance, they often rely on specialized hardware such as laser profilometers~\cite{chesser2019extrusion} or dedicated camera setups~\cite{wu2021accurate, bahrami2025optimal, fravolini2025data}, require extensive calibration iterations, and operate at the firmware level rather than through standard G-code-based printing setups. These limitations underscore the need for a more practical, accessible, and end-to-end approach for system identification and G-code generation, tailored to the needs of everyday 3D printing users.


\subsection{Contribution and Outline}

Building on the current development, this paper proposes a practical end-to-end extrusion optimization framework for high-speed FFF printing using only a phone camera for measurements and standard G-code for control implementation. The proposed method enables users to identify the extrusion and corner dynamics and optimize the extrusion control sequence in G-code in a single experimental iteration, without requiring specialized hardware or complex calibration procedures. The contributions of this paper are:
\begin{itemize}
  \item A dynamic modeling and constrained‑optimization framework that captures mis-synchronization the extrusion dynamics and cornering behavior, enabling motion–extrusion synchronization for high‑speed FFF.
  \item A practical one‑shot identification and optimization pipeline that, from two simple calibration prints to identify parameters and automatically generates optimized G‑code for most common 3D printers, facilitating high‑speed printing optimization without specialized hardware or complex calibration.
  \item A robust, hardware‑minimal phone‑camera measurement method for sub‑millimetre bead‑width estimation, supporting system identification focontrol and optimization without specialized sensors.
\end{itemize}

The remainder of this paper is organized as follows. Section~\ref{subsec:background} introduces the relevant background. Section \ref{sec:method} details the experimental setup, modeling of system and cornering, corresponding control and optimization strategies, and system identification with the phone-camera-based measurement method. Section \ref{sec:results} showcases measurement validation, identification and control experimental results and full-part printing tests. Conclusions and future directions are discussed in Section~\ref{sec:conclusion}.

\subsection{Printing dynamics formulation} \label{subsec:ExtSys}

Currently, most 3D printers execute fabrication by following G-code, a machine-neutral file generated by slicer software from CAD models. G-code defines the printhead’s motion sequence, filament feed rate, and other printing parameters.

Most slicers approximate the desired print path using linear segments. For each segment, the printhead moves a distance $D_{m}$ from its current position to a target point, while simultaneously a filament of length $D_e$ is extruded and deposited, as illustrated in Fig.~\ref{fig:3DPrintingProcess}. The extrusion ratio $\xi$ for one linear move is defined as the ratio between the extruded length and the motion distance:
\begin{equation}
  \label{eq:defExtrate}
  \xi = \frac{D_e}{D_m}\,.
\end{equation}
The extrusion ratio $\xi$ describes  the amount of material deposited per unit motion length. Over a small time interval $\mathrm{d}t$, $\xi$ can be expressed in terms of commanded velocities as:
\begin{equation}
  \label{eq:derExtrate}
  \xi = \frac{\mathrm{d}D_e}{\mathrm{d}D_m} = \frac{u_e \mathrm{d}t}{u_m \mathrm{d}t} = \frac{u_e}{u_m} ,.
\end{equation}
where $u_e$ and $u_m$ are the control command for the extrusion and the motion velocities, respectively. This shows that in a small time interval, $\xi$ represents the commanded velocity ratio , which regulates the actual velocity ratio $\zeta = v_e / v_m$. The value of $\zeta$ directly determines the deposited line width $w$ at a fixed layer height~\cite{comminal2018numerical, serdeczny2018experimental, gosset2020experimental}. Therefore, the extrusion ratio $\xi$ serves as an effective control handle to regulate the line width by adjusting the actual velocity ratio $\zeta$.  In G-code, this control is achieved by setting $D_e$ as $\xi D_m$ to enforce a desired velocity ratio and thereby produce a target width.

Conventional G-code implementations prescribe a constant extrusion ratio $\xi^{\text{c}}$, corresponding to a target constant velocity ratio $\zeta^{\text{c}}$, to produce a uniform line width $w^{\text{c}}$ in printing. While effective during constant-speed motion with $v_m^{\text{c}}$ and $v_e^{\text{c}} = \zeta^{\text{c}} v_m^{\text{c}}$, this open-loop strategy neglects the dynamic behavior of the system during transient motion phases—such as acceleration or deceleration near corners. Due to the mismatched response characteristics of the motion and extrusion subsystems, the actual velocity ratio can significantly deviate from $\zeta^{\text{c}}$, resulting in localized defects such as over- or under-extrusion.

To characterize and compensate for dynamic extrusion effects, we define spatially varying profiles along the print trajectory, parameterized by the distance $x$ from the start or from specific points (e.g., corners). These include the commanded extrusion ratio $\xi(x)$, the actual velocity ratio $\zeta(x)$, the target width $w^*(x)$, and the printed width $w(x)$. By dynamically regulating $\xi(x)$, extrusion can be better synchronized with motion, ensuring accurate velocity ratio and width control.

\section{Material and methods} \label{sec:method}

\subsection{Experimental setup}

Experiments were conducted on an Ender-3 V2 3D printer equipped with a \SI{0.4}{mm} diameter nozzle.  To enhance contrast for vision-based measurements, light ivory-colored PLA filament from Fillamentum~\cite{fillamentum} with a \SI{1.75}{mm} diameter was used for printing on a dark printing bed. We used a phone equipped with a 108-megapixels Samsung S5KHMX camera to capture images of printed curves for line width measurement. We conducted all experiments at a fixed layer height of \SI{0.2}{mm}, a bed temperature of \SI{75}{\celsius}, and a nozzle temperature of \SI{200}{\celsius}. 

This study, like the referenced works, assumes fixed process parameters such as material type, temperature, printing speed range, raster angle, infill density, and layer thickness for the specific part being produced.

\subsection{Process modeling}

To model the mis-synchronization in high-speed printing, this section first derives the extrusion dynamics model, and then analyzes cornering behavior and develop a corresponding model.

\subsubsection{Extrusion and deposition system modeling}

At a small printing layer height, the cross-section of a printed bead typically forms an elongated rectangular shape with rounded edges~\cite{comminal2018numerical,serdeczny2018experimental}. Under this bead shape, increasing the velocity ratio $\zeta$ leads to a larger cross-sectional area. Given volume conservation and a fixed layer height, this area expansion results in a  proportional increase in bead width~\cite{wu2021accurate}
Thus, the line width is approximately proportional to the velocity ratio:
\begin{equation}
  \label{eq:DerExtDepEq}
   w(x) = \alpha \cdot \zeta(x)  \, ,
\end{equation}
where the width coefficient $\alpha$ captures the linear relationship between $w$ and $\zeta$. 

To model transient dynamics, the extrusion system is commonly described using a first-order system that captures the dynamics of velocity control $u_e$ and extrusion velocity $v_e$ and the dynamics of the fast motion system can be considered as sufficiently linear~\cite{bellini2004liquefier,wu2021accurate}. The relationship between velocity ratio $\zeta$ and the G-code extrusion ratio $\xi$ solely depends on the extrusion dynamics. The corresponding transfer function is expressed as:
\begin{equation}
  \label{eq:RatioDyn}
   \frac{\zeta(s)}{\xi(s)} = \frac{1}{1+\tau(\cdot) s} \, , 
\end{equation}
where $s$ is the Laplace variable and $\tau(\cdot)$ is the system time constant. Consequently, the dynamic model between line width $w$ and the extrusion ratio $\xi$ in this work is modeled as:
\begin{equation}
  \label{eq:ExtDyn}
   \frac{w(s)}{\xi(s)} = \frac{\alpha}{1+\tau(\cdot) s} \, . 
\end{equation}
Due to materials extrusion nonlinearities, the time constant $\tau(\cdot)$  is not fixed and varies with the extrusion state~\cite{han2007coordination, chesser2019extrusion}.  To simplify the model, we approximate this nonlinearity by defining two separate time constants: $\tau_{\text{expand}}$ for increasing $\xi$, and $\tau_{\text{shrink}}$ for decreasing $\xi$, over a defined range $[\xi_{\text{low}}, \xi_{\text{high}}]$. These define the \emph{expansion} and \emph{shrinkage} models, respectively, to capture the asymmetry in the system’s transient response.

The conventional setting of $\xi(x)=\zeta^{\text{c}}$ is based on the steady-state model in  Eq.~\eqref{eq:DerExtDepEq}, where $\zeta^{\text{c}}$ satisfies $w^{\text{c}*}=\alpha \zeta^{\text{c}}$ to print desired constant $w^{\text{c}*}$. However, this static formulation neglects the dynamics  behavior by Eq.~\eqref{eq:RatioDyn}, and thus a model-based approach is required to optimize $\xi(x)$ accordingly.

\subsubsection{Corner modeling} \label{subsec:CornerModel}

\begin{figure}[h]
  \centering
  \begin{subfigure}[b]{0.44\textwidth}
  \centering
    \includegraphics[scale=1]{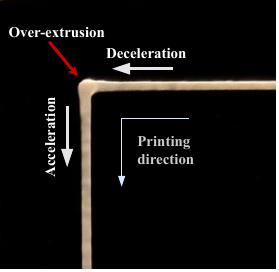} 
    \caption{Over-extrusion induced by mis-synchronization}
     \label{fig:CornerDefectImg}
  \end{subfigure}
  \hfill
  \begin{subfigure}[b]{0.55\textwidth}
  \centering
    \includegraphics[scale=1]{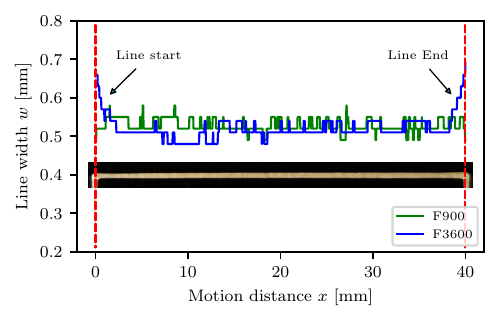} 
    \vspace{-15pt} 
    \caption{Over-extrusion width measurement}
     \label{fig:CornerDefectMeasure}
  \end{subfigure}
  \caption{Corner over-extrusion caused by deceleration mismatch at high motion velocity (\SI{3600}{\milli\meter\per\minute})}
  \label{fig:CornerDefect}
\end{figure}

Standard G-code motion commands prescribe displacements but regulate speed transitions with constant extrusion settings, resulting in passive system responses neglecting different system dynamics and leading to motion-extrusion mis-synchronization. As printing approaches a corner, sharp deceleration of $v_m(x)$ near corners outpaces the slower response of $v_e(x)$, resulting in $\zeta(x) > \zeta^{\text{c}}$ and causing over-extrusion, as illustrated in Fig.\ref{fig:CornerDefectImg}. This effect is amplified at higher speeds, where Fig.\ref{fig:CornerDefectMeasure} shows noticeable edge expansion in prints at \SI{3600}{\milli\meter\per\minute} compared to \SI{900}{\milli\meter\per\minute}.

To characterize this transient behavior quantitatively, the dynamics of the velocity ratio $\zeta(x)$ are modeled assuming that the motion system undergoes uniform acceleration or deceleration, while the extrusion system, subject to material flow inertia, responds more slowly and is considered quasi-static.

During deceleration preceding a corner, the motion velocity $v_m(x)$ decreases linearly from the constant velocity $v_m^{\text{c}}$ to zero under a constant deceleration $a$, while the extrusion velocity $v_e(x)$ remains approximately constant at $v_e^{\text{c}} = \zeta^{\text{c}} v_m^{\text{c}}$. The resulting velocity ratio and printed width along the path are:
\begin{equation}
  \label{eq:ExtrusionBeforeCorner}
   \zeta_{\text{dec}}(x) = \frac{v_{e}(x)}{v_{m}(x)} =  \frac{\zeta^{\text{c}}v_m^{\text{c}}}{\sqrt{ {v_m^{\text{c}}}^2 - 2ax}}  , 
\end{equation}
\begin{equation}
  \label{eq:WidthBeforeCorner}
   w_{\text{dec}}(x) = \alpha \zeta_{\text{dec}}(x)=  \frac{ \alpha  \zeta^{\text{c}} v_m^{\text{c}}}{\sqrt{ {v_m^{\text{c}}}^2 - 2ax}} = \frac{ v_m^{\text{c}}}{\sqrt{ {v_m^{\text{c}}}^2 - 2ax}}w^{\text{c*}}. 
\end{equation}
where $x$ denotes the spatial distance from the start of deceleration to the corner. Eq.~\eqref{eq:WidthBeforeCorner} reveals that, as approaching a corner, $v_m(x)$ drops during deceleration, and the actual width $w_{\text{dec}}(x)$ increases with the factor $\frac{ v_m^{\text{c}}}{\sqrt{ {v_m^{\text{c}}}^2 - 2ax}}$ relative to the target width $w^{\text{c}*}$, leading to width expansion at corners and causing over-extrusion due to the slower extrusion response.

Conversely, after a corner, $v_m(x)$ accelerates faster than $v_e(x)$, resulting in a temporary reduction of line width:
\begin{equation} 
\label{eq:WidthAfterCorner} 
w_{\text{acc}}(x) = \alpha  \zeta_{\text{acc}}(x) = \alpha \frac{v_e(x)}{v_m(x)} = \frac{ \alpha \zeta^{\text{c}}  v_m^{\text{c}}}{\sqrt{2ax}} = \frac{v_m^{\text{c}}}{\sqrt{2ax}} w^{\text{c}*}, \end{equation} 
where $x$ represents the spatial distance from the corner to the point where constant motion $v_m^{\text{c}}$ speed is reestablished. Although this under-extrusion is less visually apparent—partly masked by the prior over-extrusion—it must be considered in compensation design.

The transient region around a corner spans a distance:
\begin{equation} d_{\text{tr}} = \frac{{v_m^{\text{c}}}^2}{2a}, \end{equation} determined by $v_m^{\text{c}}$ and $a$. 
This modeling framework reveals how motion-extrusion mis-synchronization causes local defects, motivating the development of compensation strategies.

\subsection{Process optimization}

Building upon the preceding analysis, this section introduces a transition process compensation and optimization strategy to mitigate extrusion defects arising from motion-extrusion mis-synchronization during high-speed printing.  The approache encodes cornering behavior and system dynamics into the control implementation in G-code. Specifically, a compensated width reference $w_{\xi}^*(x)$ is designed based on the corner modeling, and a model-based optimal control is proposed to compute the optimal extrusion control sequence $\boldsymbol{\xi}^*$ for generating optimized G-code of high-speed printing.

\subsubsection{Corner transition compensation}

To mitigate the severe mis-synchronization at high-speed corner printing, a compensated width reference is first designed to compensate the dynamics of acceleration/deceleration after/before a corner in printing.

To match the fast motion deceleration before a corner, a decreasing width reference is designed in this region to enforce the extrusion deceleration in advance. The width reference in this phase is scaled according to the inverse ratio in Eq.~\eqref{eq:WidthBeforeCorner}, yielding:
\begin{equation}
  \label{eq:WScaledBeforeCorner}
    w^{*'}_{\text{dec}}(x) =  \frac{\sqrt{ {v_m^{\text{c}}}^2 - 2ax}} {v_m^{\text{c}}} w^{\text{c}*} \,.
\end{equation}
Similarly, in the acceleration region after the corner, an increasing ratio is applied to encode a in-advance rising in extrusion, aligning with the accelerating printhead motion:
\begin{equation}
  \label{eq:WScaledAfterCorner}
   w^{*'}_{\text{acc}}(x)=  \frac{\sqrt{2ax}}{v_m^{\text{c}}} w^{\text{c}*}  \,,
\end{equation}
With these two compensated design around the corner, for a straight line of length $\ell$ bounded by two corners, with a target constant width $w^{\text{c}}$, the complete compensated width reference is constructed as: 
\begin{equation}
\label{eq:wreference}
w_{\xi}^*(x) =
\begin{cases}
0, & \text{if } 0< x \leq \frac{w^{\text{c}*}}{2}, \quad \text{Stage \uppercase\expandafter{\romannumeral1}} \\
\frac{\sqrt{2ax}}{v_m^{\text{c}}} w^{\text{c}*} , & \text{if } \frac{w^{\text{c}*}}{2}< x \leq d_{\text{tr}},  \quad \text{Stage \uppercase\expandafter{\romannumeral2}} \\
w^{\text{c}*}  , & \text{if } d_{\text{tr}} \leq x \leq \ell-d_{\text{tr}},  \quad \text{Stage \uppercase\expandafter{\romannumeral3}} \\ 
\frac{\sqrt{{\left(v_m^{\text{c}}\right)}^2-2a(x-(\ell-d_{\text{tr}}))} }{{v_m^{\text{c}}}}w^{\text{c}*} , & \text{if } l-d_{\text{tr}}< x \leq \ell-\frac{w^{\text{c}*}}{2}  \quad \text{Stage \uppercase\expandafter{\romannumeral4}} \\
0, & \text{if } l-\frac{w^{\text{c}*}}{2} < x \leq \ell \quad  \text{Stage \uppercase\expandafter{\romannumeral5}} \,.
\end{cases}
\end{equation}

The compensated width reference $w_{\xi}^*(x)$ consists of the following stages:

\begin{enumerate}[label=(\roman*)]
  \item \textbf{Initial trimming} (Stage \uppercase\expandafter{\romannumeral1}): Sets the width to zero at the start to suppress over-extrusion at line junctions.

  \item \textbf{Post-corner ramp-up} (Stage \uppercase\expandafter{\romannumeral2}): Gradually increases the width after each corner to compensate for under-extrusion.

  \item \textbf{Constant-width segment} (Stage \uppercase\expandafter{\romannumeral3}): Maintains the target width for steady-state printing.

  \item \textbf{Pre-corner ramp-down} (Stage \uppercase\expandafter{\romannumeral4}): Gradually decreases the width before each corner to reduce over-extrusion.
  
  \item \textbf{Final trimming} (Stage \uppercase\expandafter{\romannumeral5}): Sets the width to zero at the end to suppress over-extrusion at line junctions.
\end{enumerate}

\begin{figure}[ht]
     \centering
     \includegraphics[scale=1]{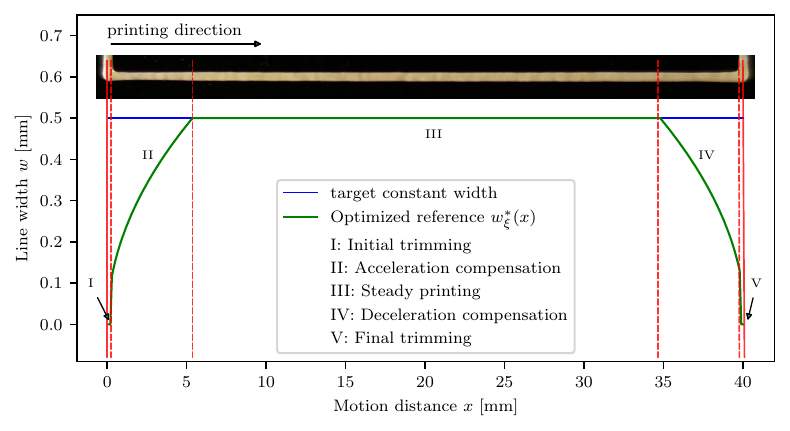}\\
     \vspace{-15pt}
     \caption{Compensating width reference for a single line}
     \label{fig:CorOptDesignExample}
\end{figure}

An example of the compensated width profile for a line with $\ell = \SI{40}{\milli\meter}$ and $w^{\text{c}*} = \SI{0.5}{\milli\meter}$ is illustrated in Fig.~\ref{fig:CorOptDesignExample}.

Extending this formulation to trajectories composed of multiple connected segments, a global compensated width reference  $w_{\xi}^*(x)$  can be constructed by applying Eq.\eqref{eq:wreference} segment-wise and concatenating the results. This reference serves as the target width reference in the model-based optimization framework detailed in the subsequent model-based optimization framework (Section\ref{sec:ExtControl}) to compute the optimal extrusion control $\boldsymbol{\xi}^*$ to generate G-code commands, improving corner transition fidelity and overall print quality at high speeds.

\subsubsection{Optimal constrained control for width tracking}\label{sec:ExtControl}

To track a target width reference, a optimal control framework is formulated based on the extrusion system dynamics introduced in Section~\ref{subsec:ExtSys}. The objective is to determine the extrusion ratio trajectory $\boldsymbol{\xi}$ that minimizes width tracking error while satisfying system constraints.

To facilitate control design, the G-code path is discretized into segments of spatial length $\Delta s$. After discretization, the originally continuous width profile $w(x)$ is represented as a sequence of discrete values $\boldsymbol{w} = \{ w_0, w_1, \dots, w_N \}$, where $w_k$ denotes the width at the $k$-th segment along the path. The first-order extrusion dynamics described in Eq.~\eqref{eq:ExtDyn} are then discretized:
\begin{equation}
  \label{eq:statefunction}
   w_{k+1} = Aw_k + B\xi_k = \left(1-\frac{\Delta s}{\tau}\right)w_k + \alpha \frac{\Delta s}{\tau}\xi_k\, ,
\end{equation}
where $\xi_k$ denote the extrusion ratio at the $k$-th spatial step, respectively. The system parameters are the width coefficient $\alpha$ and the time constant $\tau$, selected as $\tau_{\text{expand}}$ or $\tau_{\text{shrink}}$ depending on the local width transition direction, or their average for mixed cases.

Let $\boldsymbol{w^*} = \{ w^*_0, w^*_1, \dots, w^*_N \}$ be the discrete desired width sequence sampled from $w^*(x)$ at the same spatial resolution. The optimal control problem seeks to minimize the tracking error between the actual and target widths, subject to system dynamics and actuation constraints:
\begin{equation}
  \label{eq:optimization}
  \begin{aligned}
    \min_{w_{1..N}, \xi_{1..N-1}} \quad & \sum_{k=1}^{N} \big(w_k - w^*_k\big)^2 \\
    \text{subject to} \quad & w_{k+1} = Aw_k + B\xi_k, \quad for \quad k = 0..N-1 \\
                     & \xi_k \in [\xi_{low},~\xi_{high}] \quad \forall k \in \{1, 2,, \ldots, N-1\}
 \\
                     & w_0 =     \begin{cases}
      w_{\text{prev}}, & \text{if continuing from a previous segment}, \\
      0, & \text{if this is the first segment}.
    \end{cases} \,.
  \end{aligned}
\end{equation} 
The extrusion ratio bounds $[\xi_{\text{low}}, \xi_{\text{high}}]$ enforce extrusion actuator constraints, and ensure the system operates within a regime where the approximated model remains valid. The initial condition $w_0$ is defined by the final width of the preceding segment, denoted $w_{\text{prev}}$, which ensures continuity across segments and enhances print quality.

Solving this constrained optimization problem yields the optimal control sequence $\boldsymbol{\xi}^* = \{ \xi_0^*, \xi_1^*, \dots, \xi^*_N \}$, where each $\xi_k^*$ specifies the extrusion ratio for the $k$-th path segment of length $\Delta s$ with extrusion amount $\xi_k^* \Delta s$ along the discretized G-code path. For the $k$-th segment, starting at $P_k = (X_k, Y_k, Z_k)$ and ending at $P_{k+1} = (X_{k+1}, Y_{k+1}, Z_{k+1})$ with $| P_{k+1} - P_k | = \Delta s$, the G-code command is generated as:
\begin{lstlisting}
G0 X@$\{X_{k+1}\}$@ Y@$\{Y_{k+1}\}$@ Z@$\{Z_{k+1}\}$@ E@$\{\xi_k^* \Delta s\}$@
\end{lstlisting}
Note: the generated G-code specifies relative extrusion increments per segment,
\(\Delta E_k=\xi_k^*\,\Delta s\). To use absolute extrusion, replace E by its cumulative sum.

By iteratively applying this process, an optimized G-code file with spatially varying extrusion feedrates is generated, improving corner transitions and overall print quality during high-speed printing — without requiring complex hardware or firmware modifications and just need upload optimized G-code.


\subsection{End-to-end approach for G-code optimization}

\begin{figure}[ht]
     \centering     
     \includegraphics[scale=1]{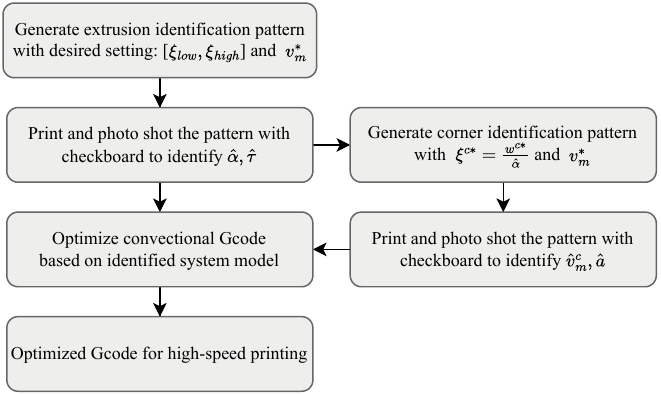}\\
     \caption{Proposed methods application flowchart}\label{fig:OverallFramework}
\end{figure}

To implement the proposed optimization, an end-to-end framework is developed that integrates system identification using only phone camera measurements to generate optimized G-code, as illustrated in Fig.~\ref{fig:OverallFramework}. 

Users simply need to specify the target printing speed $v_m^*$ and the working extrusion ratio range $[\xi_{low}, \xi_{high}]$ including the nominal working point $\xi^{c*}$ for the desired width $w^{c*}$ to start calibration and optimization:
\begin{itemize}
  \item Print a calibration pattern at the target speed $v_m^*$ to identify the width coefficient $\hat{\alpha}$ and dynamic time constants. 
  \item Using $\hat{\alpha}$, print a second pattern at $v_m^*$ to estimate cornering dynamics, specifically $\hat{v}_m^c$ and $\hat{a}$.
  \item With the identified parameters, optimize the original G-code for synchronized high-speed printing by computing the optimal extrusion ratio sequence $\boldsymbol{\xi}^*$.
\end{itemize} 
Phone cameras, as accessible vision sensors, are employed to capture images of printed calibration patterns, providing a practical and user-friendly alternative to specialized sensors~\cite{wu2021accurate, wu2023modeling, fravolini2025data} for system identification.

This approach enables users with a wide range of conventional 3D printers, to calibrate their extrusion systems and generate optimized G-code in a single experimental iteration, without specialized equipment or complex procedures. The following sections first detail the design of calibration patterns and the parameter fitting procedure, and then introduce the phone-camera-based measurement method.

\subsection{System identification for G-code optimization}

\subsubsection{Extrusion system identification}

\begin{figure}[h]
  \centering
  \begin{subfigure}[b]{0.55\textwidth}
  \centering
    \includegraphics[scale=1]{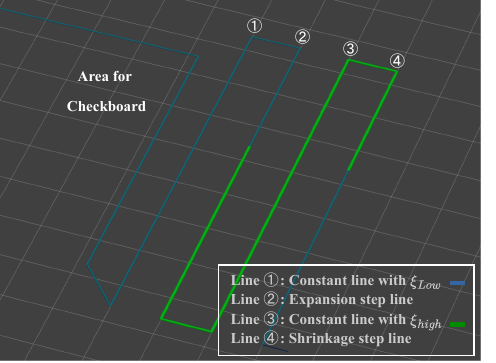} 
    \caption{G-code design}
    \label{fig:ExIDGcode}
  \end{subfigure}
  \hfill
  \begin{subfigure}[b]{0.44\textwidth}
  \centering
    \includegraphics[scale=1]{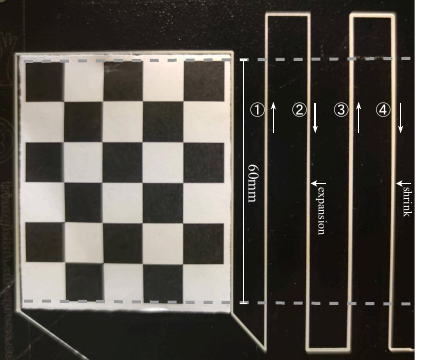} 
    \caption{Printed pattern}
    \label{fig:ExtIDPrinted}
  \end{subfigure}
  \caption{Identification pattern for extrusion system}
  \label{fig:SYSIDPattern}
\end{figure}

To optimize extrusion, a single calibration print is designed to estimate the time constants $\tau_{\text{expand}}$, $\tau_{\text{shrink}}$, and the width coefficient $\alpha$ in Eq.~\eqref{eq:statefunction}, as illustrated in Fig.\ref{fig:SYSIDPattern}. The calibration pattern consists of four lines designed to identify both steady-state and transient parameters. Measurement regions, indicated by the gray dotted lines in Fig.~\ref{fig:ExtIDPrinted}, are carefully selected to be sufficiently distant from corners to minimize the influence of corner-induced effects.

\begin{figure}[ht]
     \centering
     \includegraphics[scale=1]{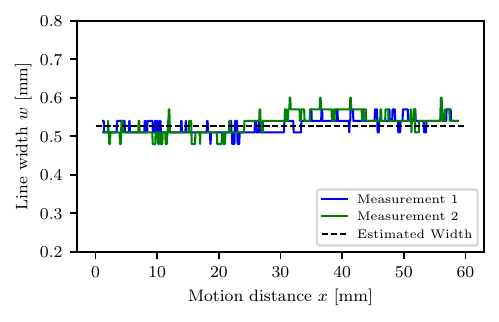}\\
      \vspace{-15pt}
     \caption{Measurements and estimation of a line of constant width lines.}
     \label{fig:measureconstantline}
\end{figure}

Lines 1 and 3 are printed with constant extrusion ratios $\xi_{low}$ and $\xi_{high}$ to estimate the width coefficient $\alpha$. As shown in Fig.~\ref{fig:measureconstantline}, repeated measurements of constant-width lines exhibit consistent and centralized distributions with noise. To estimate the constant width, the measurements are modeled as a Gaussian distribution:
\begin{equation}
  \label{eq:WidthDistribution}
   W_{\text{mes}} \sim \mathcal{N}(W_{\mu}, \sigma_{\text{mes}}^2) \,,
\end{equation}
where $W_{\text{mes}}$ represents the measured width, $W_{\mu}$ the mean, and $\sigma_{\text{mes}}$ the standard deviation of measurement noise. The Maximum Likelihood Estimator (MLE) is used to compute the estimated widths $\hat{W}_{\text{MLE,high}}$ and $\hat{W}_{\text{MLE,low}}$. The width coefficient $\hat{\alpha}$ is then calculated as:
\begin{equation}
    \hat{\alpha} = \frac{1}{2} \left( \frac{\hat{W}_{\text{MLE,high}}}{\xi_{high}} + \frac{\hat{W}_{\text{MLE,low}}}{\xi_{low}} \right) \,.   
\end{equation}

\begin{figure}[ht]
  \centering
  \begin{subfigure}[b]{0.49\textwidth}
    \includegraphics[scale=1]{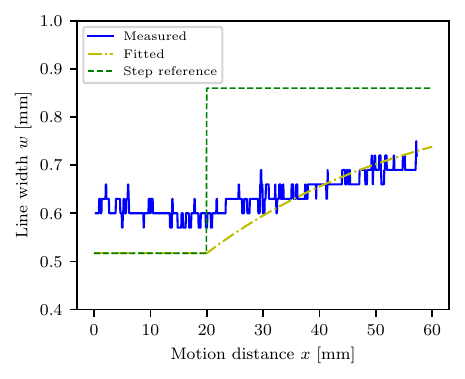} 
    \vspace{-15pt}
    \caption{Curves of expansion model identification}
    \label{fig:ExpandSysID}
  \end{subfigure}
  \hfill
  \begin{subfigure}[b]{0.49\textwidth}
    \includegraphics[scale=1]{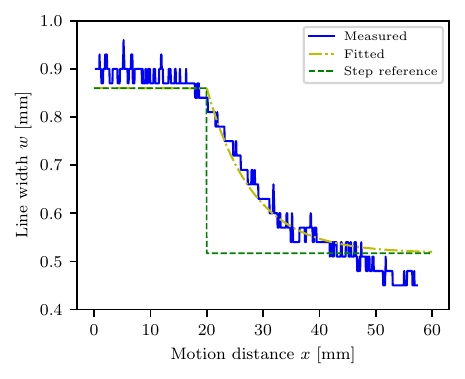} 
    \vspace{-15pt}  
    \caption{Curves of shrinkage model identification}
    \label{fig:ShrinkSysID}
  \end{subfigure}
  \caption{System identification of transient models}
  \label{fig:TranSysID}
\end{figure}

Lines 2 and 4 feature step changes in $\xi(x)$ between $\xi_{\text{low}}$ and $\xi_{\text{high}}$ to capture transient dynamics, with transition points marked in Fig.~\ref{fig:ExtIDPrinted}. The time constants $\tau_{\text{expand}}$ and $\tau_{\text{shrink}}$ are extracted by fitting the measured width response to a first-order model:
\begin{equation}
w(x) = W^{-} + (W^{+} - W^{-}) \cdot \left(1 - e^{-\frac{x}{\tau}}\right),
\label{eq:ExtFirst}
\end{equation}
where $W^{-}$ and $W^{+}$ represent pre- and post-transition steady-state widths, derived as $\hat{\alpha} \xi_{\text{low}}$ and $\hat{\alpha} \xi_{\text{high}}$ using the previously identified $\hat{\alpha}$. For the step-up transition (Line 2), $(W^{-}, W^{+}) = (\hat{W}_{\text{MLE,low}}, \hat{W}_{\text{MLE,high}})$, and vice versa for the step-down transition (Line 4). Fig.~\ref{fig:TranSysID} shows the measured width profiles and fitted curves for both expansion and shrinkage transitions.

\subsubsection{High-speed corner printing identification}

\begin{figure}[ht]
  \centering
  \begin{subfigure}[b]{0.55\textwidth}
  \centering
    \includegraphics[scale=1]{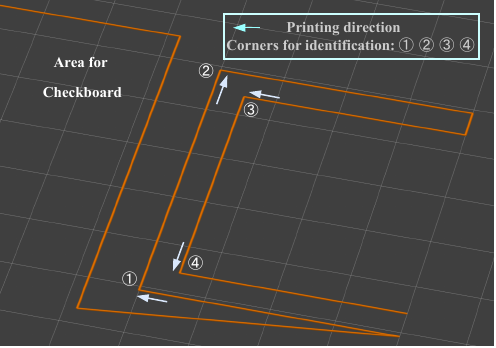} 
    \caption{G-code design}
    \label{fig:CornerIDGcode}
  \end{subfigure}
  \hfill
  \begin{subfigure}[b]{0.44\textwidth}
   \centering
    \includegraphics[scale=1]{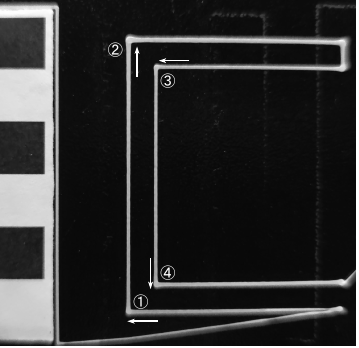}
    \caption{Printed pattern}
    \label{fig:CornerIDPrinted}
  \end{subfigure}
  \caption{Identification pattern for corner modeling}
  \label{fig:CornerIDPattern}
\end{figure}

To enable effective corner transition compensation, the parameters $a$ and $v_m^{\text{c}}$ are identified using a high-speed corner calibration pattern (Fig.~\ref{fig:CornerIDPattern}) featuring four isolated corners. The pattern is printed with a constant extrusion ratio $\xi(x) = \zeta^c$, where $\zeta^c$ is chosen to achieve the target width $w^{c*}$ based on the relationship $\hat{\alpha} \zeta^c = w^{c*}$.

\begin{figure}[ht]
  \centering
  \begin{subfigure}[b]{0.49\textwidth}
  \centering
    \includegraphics[scale=1]{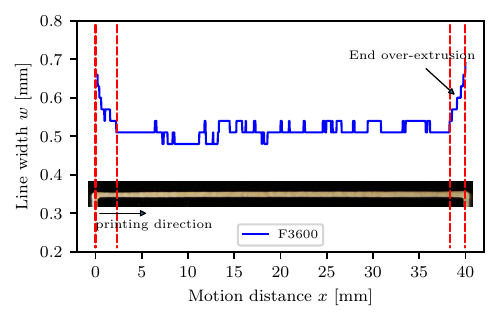} 
    \vspace{-25pt} 
    \caption{Corner measure example}
    \label{fig:CornerIDMeasure}
  \end{subfigure}
  \hfill
  \begin{subfigure}[b]{0.49\textwidth}
   \centering
    \includegraphics[scale=1]{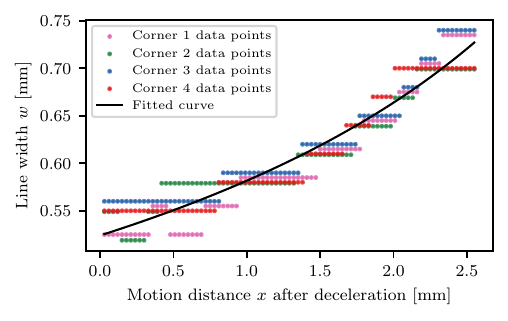} 
    \vspace{-25pt} 
    \caption{Corner parameters fitting}
    \label{fig:CornerIDFitdemo}
  \end{subfigure}
  \caption{Measure and fitting from corner identification pattern with at \SI{3600}{\milli\meter\per\minute} .The identified parameters are $\hat{a}= \SI{406}{mm/s^2}$ and $\hat{v}_{0}= \SI{66}{mm/s}$.}
  \label{fig:CornerIDFit}
\end{figure}

As shown in Fig.~\ref{fig:CornerIDMeasure}, the width measurement of a line with corners at both ends reveals over-extrusion at the corners, while the middle segment achieves the target width $w^{c*}$. For parameter identification, the deceleration-induced over-extrusion at the end of the line is analyzed. Fig.~\ref{fig:CornerIDFitdemo} illustrates the fitting of Eq.~\eqref{eq:WidthBeforeCorner} to the deceleration regions of four corners, yielding the identified parameters $\hat{v}_m^{\text{c}}$ and $\hat{a}$. These fitted parameters, derived under the proposed corner modeling assumptions, may exhibit deviations from nominal printing settings.

The deceleration region is manually segmented at the point where the line width begins to increase. Measurements are adjusted to ensure consistent pre-deceleration widths across corners, mitigating the effects of uneven leveling or measurement noise. 

\subsection{Phone-Camera-Based line width Measurement}
\begin{figure}[ht]
     \centering
     \includegraphics[scale=1]{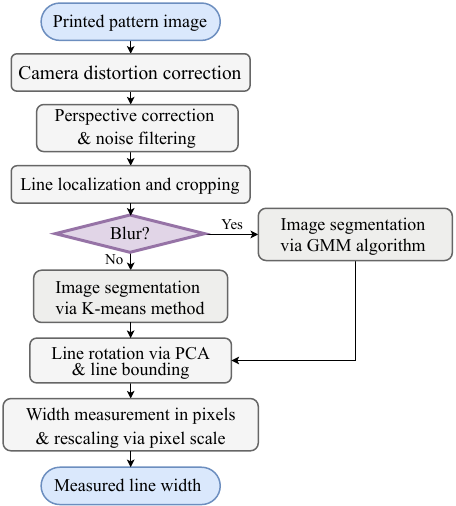}\\
     \caption{Line width measurement flowchart}
     \label{fig:VisionMeasureFlowchart}
\end{figure}
Phone cameras offer high accessibility and resolution, yet achieving precise sub-millimeter line width measurements remains challenging due to factors such as noise, blurring, perspective and lens distortions, and unknown pixel-to-metric scaling. To address these challenges, we propose a robust framework integrating advanced image processing and unsupervised clustering techniques, enabling accurate bead width measurements using phone cameras.

Fig.~\ref{fig:VisionMeasureFlowchart} outlines the proposed measurement workflow. The process begins with perspective and lens distortion correction, followed by denoising and region-of-interest extraction. Unsupervised clustering algorithms, K-means~\cite{lloyd1982,hastie2009} or Gaussian Mixture Models (GMM)~\cite{dempster1977,bishop2006}, are applied for robust bead segmentation based on image quality. Principal Component Analysis (PCA) determines the bead orientation, enabling width measurement orthogonal to the print path. Finally, pixel measurements are converted to physical units using the calibrated pixel scale.



The following sections detail the key steps: perspective correction and image segmentation.

\subsubsection{Perspective Correction and Resizing}

\begin{figure}[ht]
    \centering
    \begin{subfigure}[t]{0.4\textwidth}
        \centering
        \includegraphics[height=0.6\textwidth]{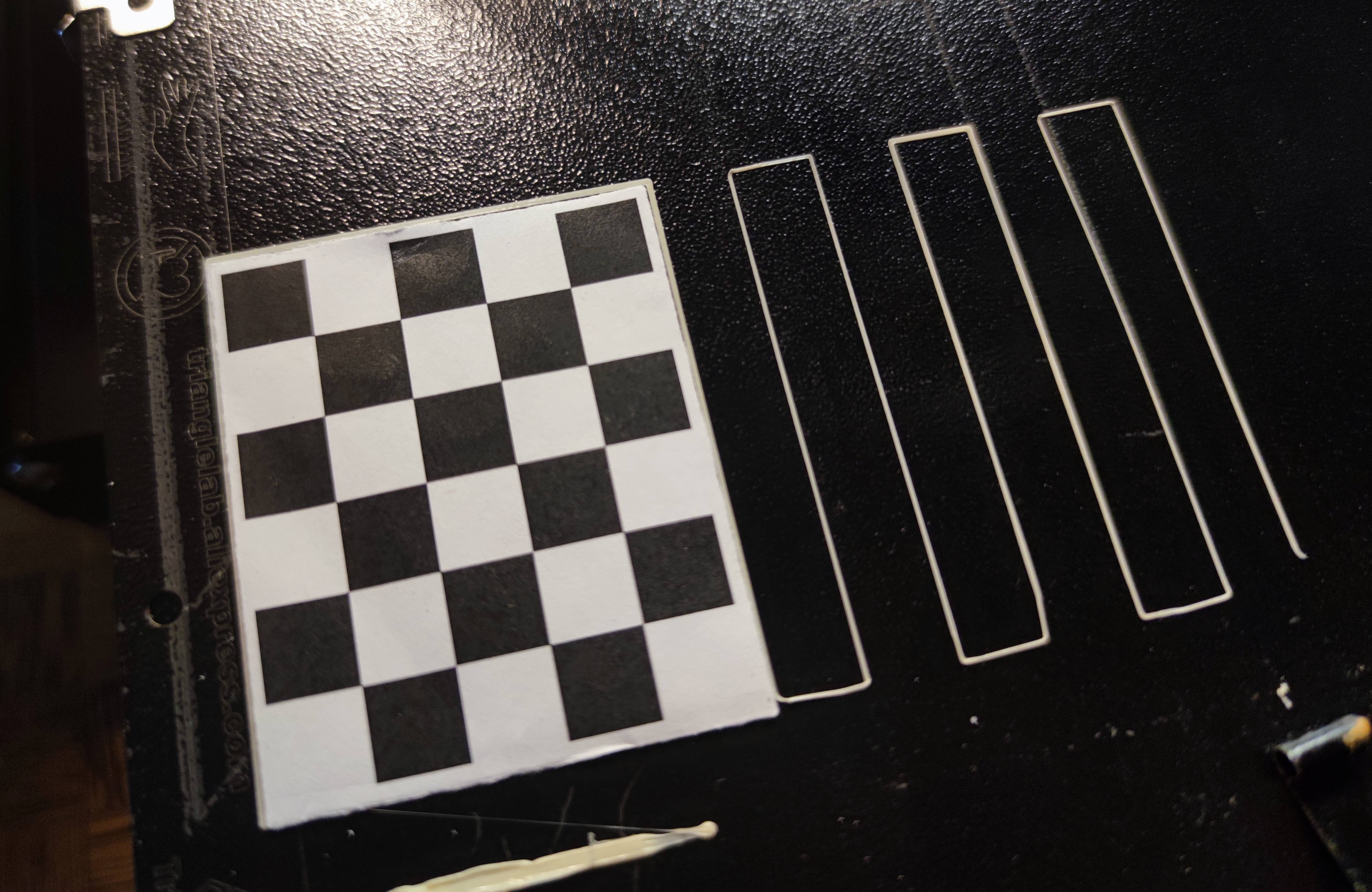}
        \caption{Printed lines and checkerboard pattern on the printer bed}
        \label{fig:PerspectiveDistortion}
    \end{subfigure} \hspace{0.025\textwidth} 
    \begin{subfigure}[t]{0.4\textwidth}
        \centering
        \includegraphics[height=0.6\textwidth]{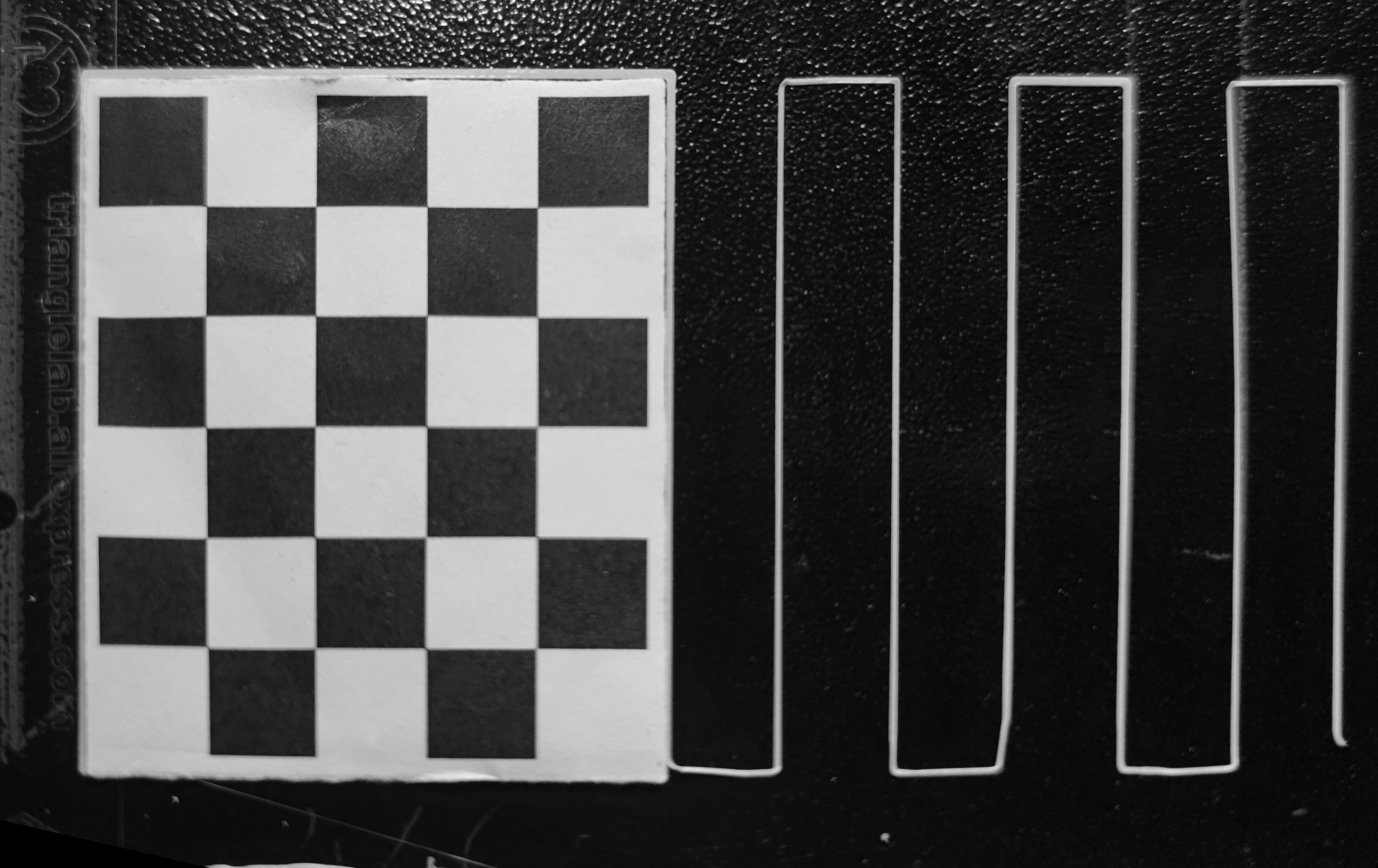}
        \caption{Image after Perspective Correction and Resizing}
        \label{fig:PerspectiveCorrect}
    \end{subfigure}

    \vspace{0.15cm} 
    \begin{subfigure}{0.8\textwidth}
        \centering
        \includegraphics[width=\textwidth]{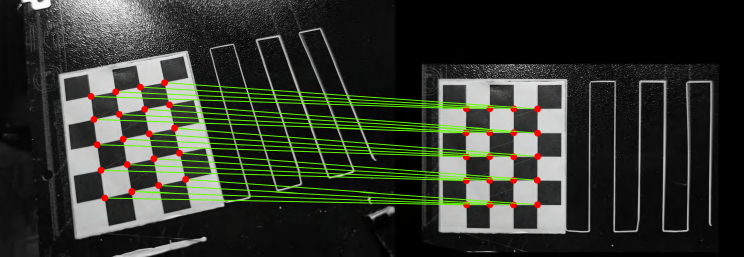}
        \caption{Homograph projection based on the checkerboard pattern}
        \label{fig:PerspectiveMap}
    \end{subfigure}
    
    \caption{Perspective Correction and Resizing}
    \label{fig:Perspective Correction and Resizing}
\end{figure}

Perspective distortion from by non-ideal camera alignment, introduces spatially varying pixel-to-metric scaling, complicating accurate width measurements (Fig.~\ref{fig:PerspectiveDistortion}). To address this, we use a printed checkerboard calibration pattern on the printer bed for perspective correction and pixel scaling, enabling flexible camera positioning with accessible checkerboard-based calibration methods, instead of specialized camera setups~\cite{wu2021accurate,fravolini2025data}.


As shown in Figure~\ref{fig:PerspectiveDistortion}, the distorted image $I_{\text{distorted}}$ is rectified using the checkerboard pattern as a reference. Assuming the checkerboard lies flat on the printing surface with known geometric specifications, the homography matrix $\mathcal{H}$ is computed by solving the transformation:
\begin{equation}
  \mathcal{P}_{\text{corrected}} = \mathcal{H} \cdot \mathcal{P}_{\text{distorted}},
\end{equation}
where $\mathcal{P}_{\text{distorted}}$ represents the detected corner points in the checkerboard in $I_{\text{distorted}}$, and $\mathcal{P}_{\text{corrected}}$ denotes their corresponding ideal positions in the rectified image $I_{\text{corrected}}$, as shown in Figure~\ref{fig:PerspectiveMap}. $\mathcal{H}$ is estimated by minimizing the reprojection error between $\mathcal{P}_{\text{distorted}}$ and $\mathcal{P}_{\text{corrected}}$, and $I_{\text{distorted}}$ can then be corrected by homography transformation: 
\begin{equation}
  \mathcal{I}_{\text{corrected}} = \mathcal{H} \cdot \mathcal{I}_{\text{distorted}}.
\end{equation}
The corrected image $I_{\text{corrected}}$, as shown in Fig.~\ref{fig:PerspectiveCorrect}, is in a top-down perspective with uniform pixel-to-metric scaling. The pixel-to-metric scale is determined from the known physical dimensions and corresponding pixel distances of the checkerboard pattern.

\subsubsection{Printed Lines Segmentation}
\begin{figure}[h]
     \centering     
     \includegraphics[scale=1]{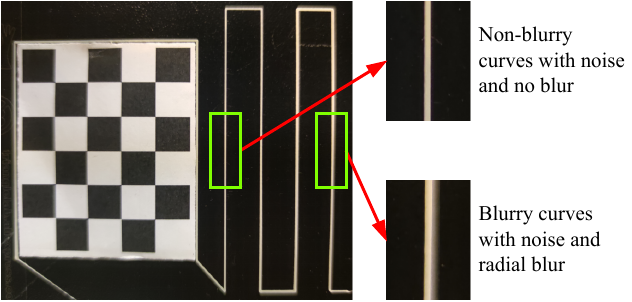}\\
     \caption{Details of photographied printed lines}\label{fig:ImageSegment}
\end{figure}

As shown in Fig.~\ref{fig:ImageSegment}, central lines appear sharp but are affected by background noise, while off-center lines suffer from radial blur due to the lens's shallow depth of field. To address this, we employ unsupervised clustering methods to robustly mitigate noise and blur effects to improve segmentation accuracy for width measurements.

\begin{figure}[h]
  \centering
  \begin{subfigure}[b]{0.49\textwidth}
    \includegraphics[width=\linewidth]{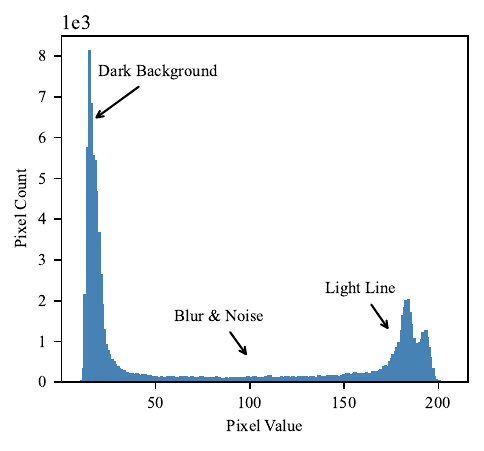}
    \vspace{-30pt}
    \caption{Non-blurry line}
    \label{fig:DistClear}
  \end{subfigure}
  \hfill
  \begin{subfigure}[b]{0.49\textwidth}
    \includegraphics[width=\linewidth]{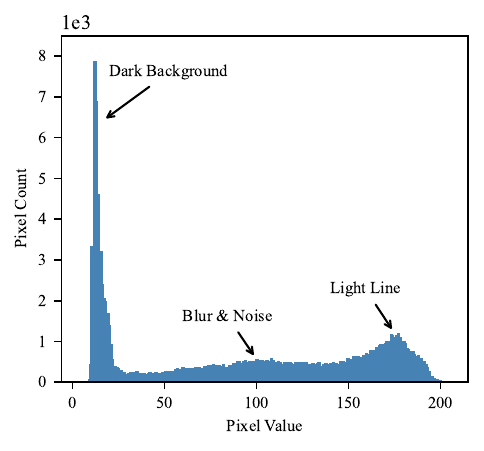}
    \vspace{-30pt}  
    \caption{Blurry line}
    \label{fig:DistArtifacts}
  \end{subfigure}
  \caption{Pixel value histograms}
  \label{fig:distribution}
\end{figure}


The pixel value histograms in Fig.~\ref{fig:distribution} reveal distinct clusters corresponding to different regions for both blurry and non-blurry lines, and these pixels can be culstered with unsupervised clustering methods by assign each pixel to a cluster based on its intensity. The relevant pixel groups for segmentation are categorized as follows:
\begin{itemize}
    \item{\textit{Dark background}, $\mu_{dark}, \sigma_{dark}$: Pixels of the black background, with low pixel value.}
    \item{\textit{Light line}, $\mu_{light}, \sigma_{light}$: Pixel of the white printed lines, with high pixel value.}
    \item{\textit{Noise or blur}, $\mu_{noise}, \sigma_{noise}$: Pixels with values between the previous two groups.}
\end{itemize}
Each cluster is characterized by a centroid $\mu$ and a variance $\sigma$, representing its pixel value distribution. We employ both K-means and GMM clustering for pixel classification into three groups. 

K-means clusters pixels by minimizing intra-cluster variance. Following K-means clustering, the segmentation threshold is defined as:
\begin{equation}
  \label{eq:Threshold}
   \text{Threshold}_{K\text{-means}} = \mu_{\text{noise}} + n \cdot \sigma_{\text{noise}} \,,
\end{equation}
where $n \in [1, 2.5]$ is an adjustable parameter. Pixels with intensities above this threshold are classified as part of the line.

GMM models pixel intensity distributions as a mixture of Gaussians, with parameters estimated using the Expectation-Maximization algorithm. The cluster with the highest mean intensity is identified as the printed line.

Based on validation results (Sec.~\ref{sec:MeasureValidate}), K-means is applied to non-blurry images, while GMM is used for blurry regions.


\section{Calculation and Resultss}  \label{sec:results}
\subsection{Phone-based measurements validation} \label{sec:MeasureValidate}

The proposed phone-camera-based measurement methods were validated by comparing their results against ground truth data obtained using a high-resolution KEYENCE LM-X Series microscope~\cite{keyence_lmx}. Measurements were conducted on 4 clear and 4 blurry lines to evaluate the accuracy and robustness of the proposed approach under varying image quality conditions. 

\begin{figure}[h]
  \centering
  \begin{subfigure}[b]{0.49\textwidth}
    \includegraphics[width=\linewidth]{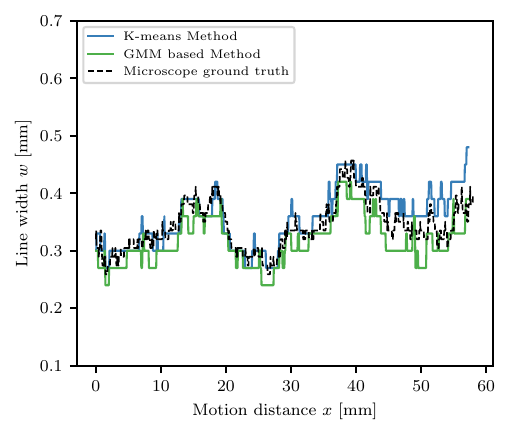}
    \vspace{-30pt}
    \caption{Clear line measurements validation}
    \label{fig:ValidationClear}
  \end{subfigure}
  \begin{subfigure}[b]{0.49\textwidth}    
      \includegraphics[width=\linewidth]{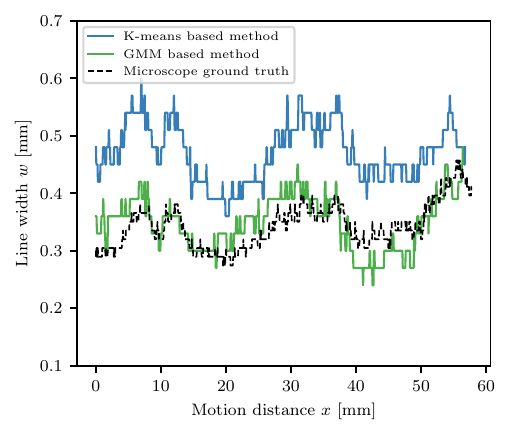}    \vspace{-30pt}
      \caption{Blurry line measurements validation}
      \label{fig:ValidationBlur}
      \end{subfigure}
  \caption{Examples of validation of phone-camera-based measurements}
  \label{fig:GroundValidation}
\end{figure}

Figure~\ref{fig:GroundValidation} summarizes the validation results. For non‑blurry images (Fig.~\ref{fig:ValidationClear}), both methods closely match the microscope ground truth. For blurry images (Fig.~\ref{fig:ValidationBlur}), the GMM method remains robust and aligns well with the ground truth, whereas the K‑means method systematically overestimates widths due to misclassification of blurred pixels.

RMSEs and MAPEs between the microscope ground truth and the proposed phone‑camera methods are summarized in Table~\ref{tab:method_comparison}. For non‑blurry images, the K‑means method attains an RMSE of \SI{0.044}{\milli\meter} and a MAPE of 7.4\%, but its accuracy degrades on blurred regions. The GMM method is more robust to blur, producing an RMSE of \SI{0.069}{\milli\meter} and a MAPE of 12.4\% overall. Therefore, we apply K‑means on sharp image areas and GMM on blurred areas to balance measurement accuracy and robustness.

\begin{table}[htbp]
\centering
\begin{tabular}{*{4}{c}}  
\toprule
\textbf{Scene} & \textbf{Metric} & \textbf{K-means Method} & \textbf{GMM Method} \\ 
\midrule
\multirow{2}{*}{Non-blurry} 
& RMSE [\SI{}{\milli\meter}] & 0.044 & 0.073 \\ 
& MAPE [\%] & 7.4 & 11.9 \\ \hline
\addlinespace
\multirow{2}{*}{Blurry} 
& RMSE [\SI{}{\milli\meter}] & 0.113 & 0.069 \\ 
& MAPE [\%] & 22.4 & 12.4 \\ 
\bottomrule
\end{tabular}
\caption{Performance comparison between K-means and GMM based methods for line measurement accuracy. Root Mean Square Error (RMSE) values are in millimeters (\SI{}{\milli\meter}), and  Mean Absolute Percentage Error (MAPE) values are percentages. Ground truth obtained by microscope.}
\label{tab:method_comparison}
\end{table}

These results show the phone‑camera measurement yields sufficiently accurate width estimates for reliable one‑shot system identification and control optimization. The residual error versus the microscope is on the order of 1–2 camera pixels, negligible for the intended tasks.

\subsection{Extrusion system identification and optimal control}

This section presents the results of extrusion system identification using the proposed one-shot calibration method and evaluates the optimal control strategy for width tracking. The calibration pattern, designed for an extrusion ratio range of $0.03$–$0.05$, validated the proposed identification and control's effectiveness in improving extrusion synchronization and print quality.

\subsubsection{Extrusion Identification}



The calibration pattern was designed with $\xi_{high} = 0.05$ and $\xi_{low} = 0.03$, as illustrated in Fig.~\ref{fig:TranSysID}. System identification was conducted at \SI{900}{\milli\meter\per\minute} and \SI{3600}{\milli\meter\per\minute} with repeated experiments were performed using the same one-shot calibration pattern. The identified steady-state width coefficient $\hat{\alpha}$ and transient time constants $\hat{\tau}_{expand}$ and $\hat{\tau}_{shrink}$ are summarized in Table~\ref{tab:SysIDResult}.


\begin{table}[h]
\centering
\begin{tabular}{cccccc}
\hline
Speed (mm/min) &  $\hat{\alpha}$ & $\hat{\tau}_{expand}$ & $\hat{\tau}_{shrink}$ \\ \hline
900   &  16.98 $\pm$ 0.36 & 37.81 $\pm$ 1.17 & 8.80 $\pm$ 0.82 \\ \hline
3600  &  16.67 $\pm$ 0.18 & 37.04 $\pm$ 0.58 & 9.98 $\pm$ 2.57 \\ \hline
\end{tabular}
\caption{Mean and standard deviation of identified extrusion system parameters at two printing speeds. Each value is obtained from two repeated experiments using the proposed one-shot calibration pattern. The small variation confirms the robustness and repeatability of the identification method.}
\label{tab:SysIDResult}
\end{table}
Minor variations across repetitions are attributed to measurement noise and slight fluctuations in printing conditions, such as extrusion pressure or temperature. These deviations are relatively minimal, demonstrating the robustness and repeatability of the proposed identification method for subsequent optimal control.

The consistent values of $\hat{\alpha}$, $\hat{\tau}_{expand}$, and $\hat{\tau}_{shrink}$ across speeds indicate that the extrusion system dynamics are largely invariant within the tested range, validating the applicability of low-speed calibration for high-speed G-code optimization.

\subsubsection{Optimal constrained control for desired line width tracking}

Using the identified extrusion system parameters, optimal constrained control is applied to generate the optimal extrusion ratio sequence $\boldsymbol{\xi}^*$ for width reference tracking. To track desired reference with width between \SI{0.524}{mm} ($\xi_{low} = 0.03$) and \SI{0.854}{mm} ($\xi_{high}  = 0.05$), the printing path is discretized with $\Delta s = \SI{0.1}{mm}$ and the extrusion ratio is constrained within $[-2, 2]$ to limit extruder jerk. 

\begin{figure}[h]
  \centering
  \begin{subfigure}[b]{0.495\textwidth}
    \centering
    \includegraphics[scale=0.92]{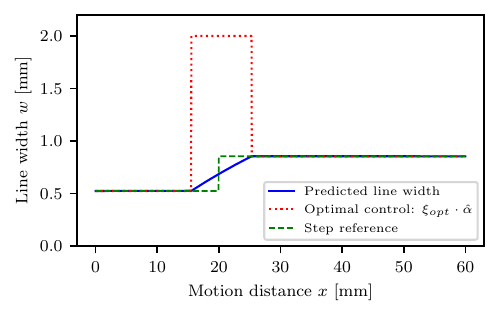}
    \vspace{-15pt} 
    \subcaption{Optimal control for width expansion step}
    \label{fig:ExpandOptCtrl}
  \end{subfigure}
  \hfill
  \begin{subfigure}[b]{0.495\textwidth}
    \centering
    \includegraphics[scale=0.92]{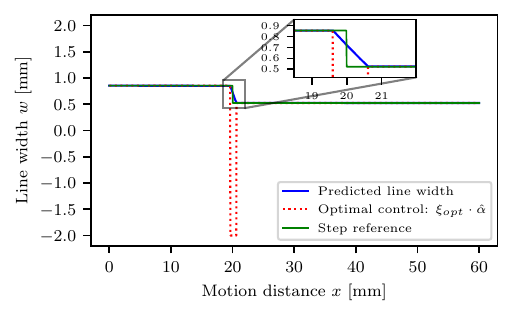}
    \vspace{-15pt} 
    \subcaption{Optimal Control for width shrinkage step}
    \label{fig:ShrinkOptCtrl}
  \end{subfigure}
  \caption{Optimal constrained control for desired width tracking.  $\boldsymbol{\xi}^*$ is scaled by $\hat{\alpha}$ for clearer visualization, enabling comparison with the reference and predicted line widths.}
  \label{fig:OptCtrlSim}
\end{figure}

Solving the optimization problem in Eq.~\eqref{eq:optimization}, the optimal control sequence is shown in
Fig.~\ref{fig:OptCtrlSim}. It can be observed that, to compensate the relative slow response of the extruion system, the optimal control $\boldsymbol{\xi}^*$ reaches the upper / lower bound in advance of the expansion / shrinkage step change to track the desired width reference.

\begin{figure}[h]
  \centering
  \begin{subfigure}[b]{\textwidth}
    \centering
    \includegraphics[scale=0.9]{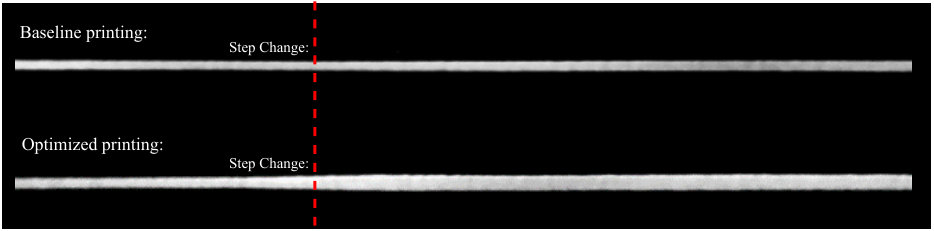}
    \vspace{-5pt} 
    \caption{Printed lines during width expansion step}
    \label{fig:PrintedExpandCtrl}
  \end{subfigure}
  \begin{subfigure}[b]{\textwidth}
    \centering
    \includegraphics[scale=0.9]{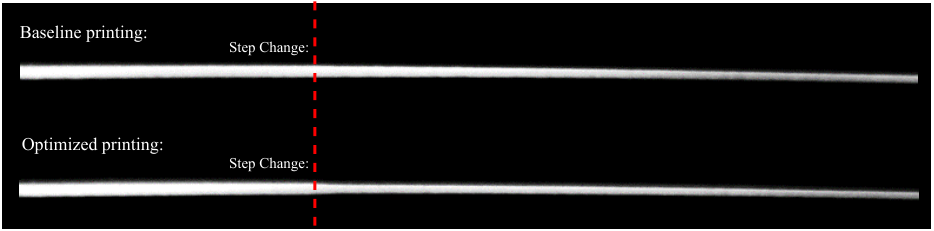}
    \vspace{-5pt} 
    \caption{Printed lines during width shrinkage step}
    \label{fig:PrintedShrinkCtrl}
  \end{subfigure}
  \caption{Comparison of width steps obtained with the baseline approach and with optimal control. Step change red dash lines indicate the location of expansion/shrinkage step change in reference.}
  \label{fig:PrintedOptStep}
\end{figure}

 Fig.~\ref{fig:PrintedOptStep} shows the lines printed with the proposed control alongside with the baseline lines produced, where the extrusion control input directly corresponds to the desired line width step. The optimized lines exhibit faster and more precise transitions in both expansion and shrinkage steps. 

\begin{figure}[h]
  \centering
  \begin{subfigure}[b]{0.49\textwidth}
    \centering
    \includegraphics[scale=1]{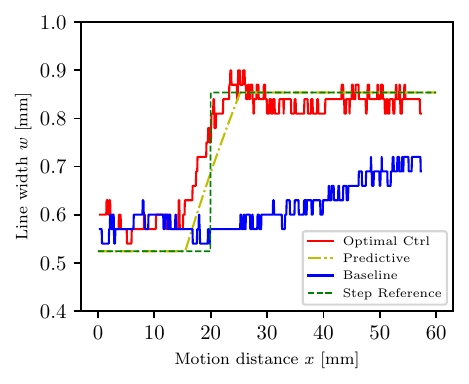}
    \vspace{-15pt} 
    \caption{Measurement of expansion step tracking}
    \label{fig:ExpandCtrl}
  \end{subfigure}
  \hfill
  \begin{subfigure}[b]{0.49\textwidth}
    \centering
    \includegraphics[scale=1]{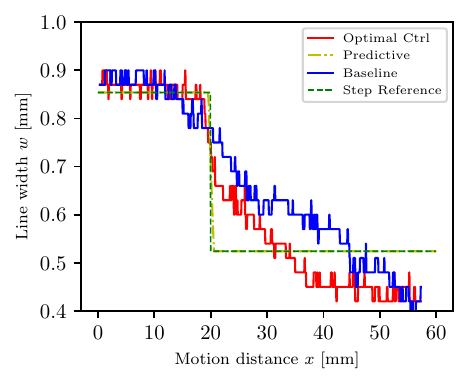}
    \vspace{-15pt} 
    \caption{Measurement of shrinkage step tracking}
    \label{fig:ShrinkCtrl}
  \end{subfigure}
  \caption{Measurement of optimal constrained control for width tracking}
  \label{fig:OptStep}
\end{figure}

Table \ref{tab:RMSEctrl} summarizes the RMSEs between the desired width reference and the printed line widths, highlighting the improvements achieved with the proposed optimal control. The method effectively reduces the tracking RMSE for both expansion and shrinkage steps to below \SI{0.06}{\milli\meter}, with a notable 69.95\% improvement in the expansion step.

\begin{table}[h]
    \centering
    \begin{tabular}{cccc}
    \hline
                & RMSE of baseline [\SI{}{\milli\meter}] & RMSE of optimal control [\SI{}{\milli\meter}] & Improvement \\ \hline
    Expansion step & 0.166                       & 0.050                        & 69.95\%      \\ 
    Shrinkage step & 0.070                       & 0.059                        & 15.61\%      \\ \hline
    \end{tabular}
  \caption{RMSE of baseline and optimal control cases with respect to the desired width reference}
  \label{tab:RMSEctrl}
\end{table}

\subsection{Corner identification and compensation}

\begin{figure}[ht]
    \begin{subfigure}[b]{0.4\textwidth}
         \includegraphics[scale=1]{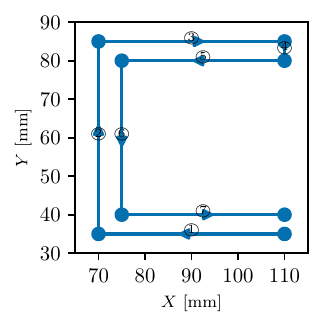}\\
             \vspace{-30pt} 
         \caption{Cornering cornering pattern to be compensated}
         \label{fig:CornerPatternSimple}
    \end{subfigure}
    \hfill
     \begin{subfigure}[b]{0.6\textwidth}
        \includegraphics[scale=1]{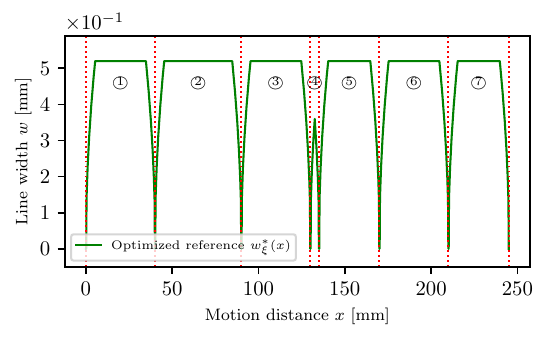}\\    
         \vspace{-30pt} 
         \caption{Compensation width reference for the cornering calibration pattern}
         \label{fig:CaliCorOptWidthReference}
    \end{subfigure}
    
    \begin{subfigure}[b]{\textwidth}
     \centering
         \includegraphics[scale=1]{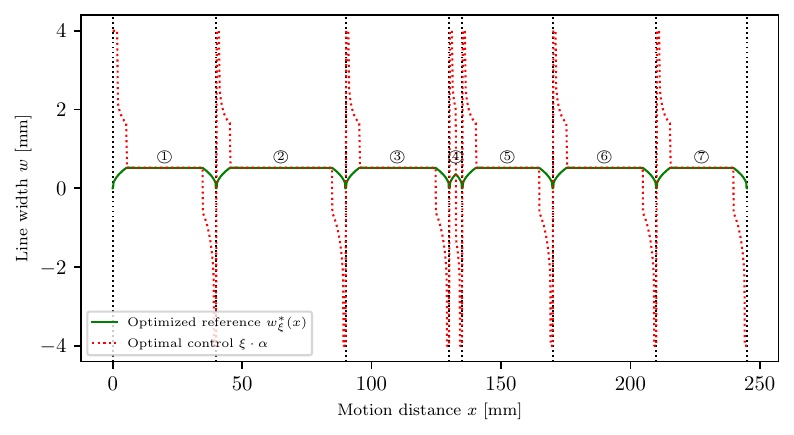}\\    
         \vspace{-10pt} 
         \caption{Optimized constrained control for $w_{\xi}^*(x)$ tracking, for the cornering calibration pattern}
         \label{fig:CaliCorOptCtrl}
    \end{subfigure}
    \caption{Cornering calibration pattern after corner defects compensation reference and control}
    \label{fig:CornerCompensationExample}
\end{figure}

By printed a corner calbration pattern shown in Fig.~\ref{fig:CornerIDPattern} at  \SI{3600}{\milli\meter\per\minute}, the cornering parameters are identified as shown in Fig.~\ref{fig:CornerIDFit} with $\hat{a}= \SI{406}{mm/s^2}$ and $\hat{v}_{0}= \SI{66}{mm/s}$. Using the identified parameters of extrusion and corner, high-speed cornering defects can be compensated with proposed methods as shown in Fig.~\ref{fig:CornerCompensationExample}.

To address high-speed corner printing defects for the pattern in Fig.~\ref{fig:CornerPatternSimple}, a compensating width reference $w_{\xi}^*(x)$ is designed, as shown in Fig.~\ref{fig:CaliCorOptWidthReference}. By applying the proposed optimal constrained control to track this reference, the resulting optimized extrusion input $\boldsymbol{\xi}^*$ is depicted in Fig.~\ref{fig:CaliCorOptCtrl}.


\begin{figure}[ht]
  \centering
  \begin{subfigure}[b]{0.23\textwidth}
    \includegraphics[width=\textwidth]{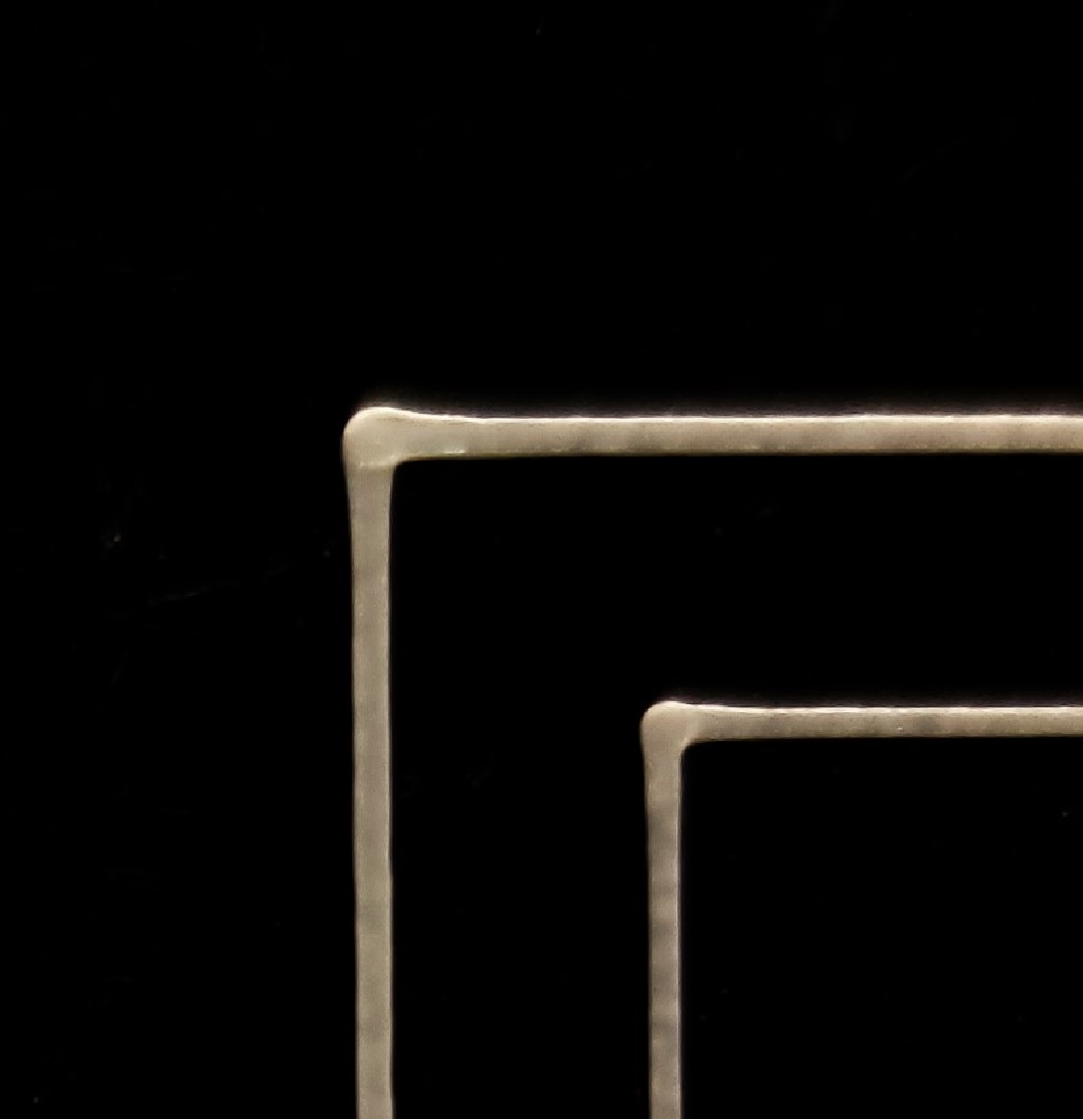}
    \caption{Baseline Corner Printing}
    \label{fig:Corner3600NoOpt}
  \end{subfigure}
  \hspace{1cm}
  \begin{subfigure}[b]{0.23\textwidth}
    \includegraphics[width=\textwidth]{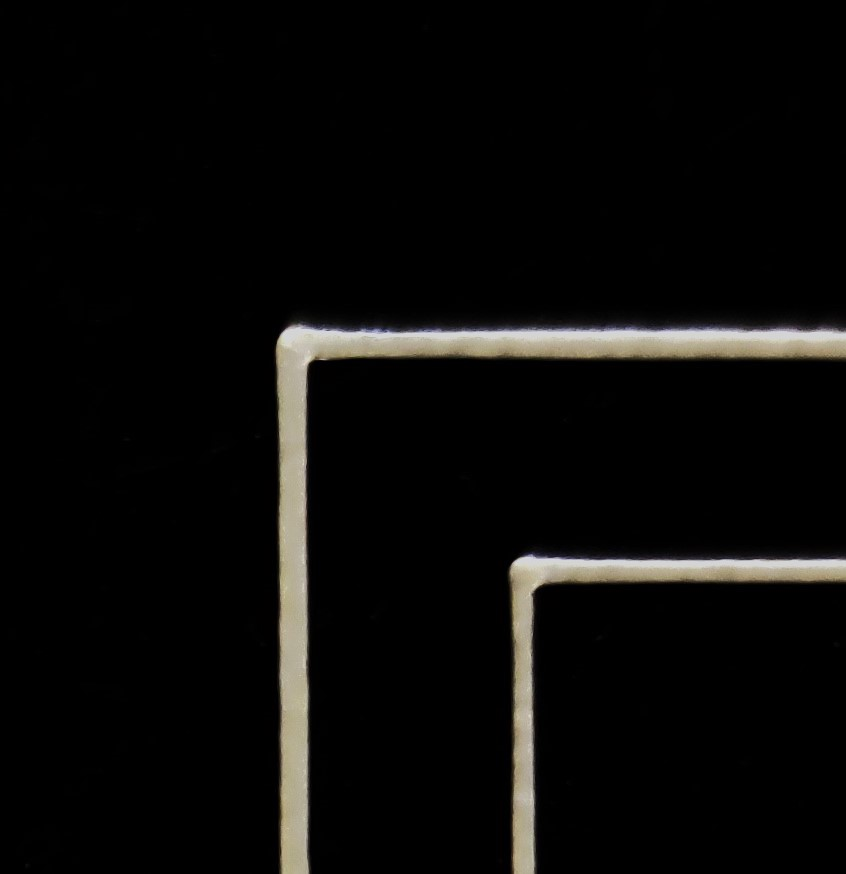}
    \caption{Optimized Corner Printing}
    \label{fig:Corner3600Opt}
  \end{subfigure}
    \vspace{-10pt} 
  \caption{High-speed corner printing comparison}
  \label{fig:CornerOptCompare}
\end{figure}

\begin{figure}[ht]
     \centering
     \includegraphics[scale=1]{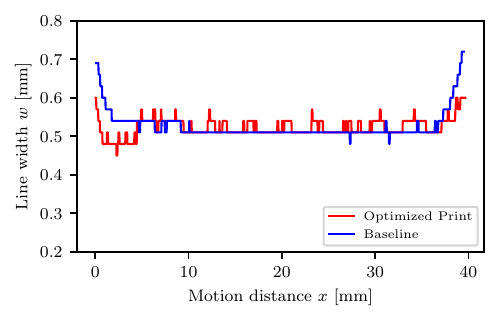}\\
    \vspace{-10pt} 
     \caption{Measurements of high-speed printed corners with and without optimization}
     \label{fig:CorOpt}
\end{figure}

Figure~\ref{fig:CornerOptCompare} compares high-speed printed corners with and without optimization, while Figure~\ref{fig:CorOpt} presents the corresponding line width measurements. The images and measurements demonstrate that the proposed optimization effectively mitigates corner defects.

Table~\ref{tab:CornerQA} provides a quantitative analysis of the optimization. The maximum deviation between the measured and desired line widths, as well as the variance of the measurements, are reported to assess the quality and consistency of the deposited lines. The proposed method reduces the maximum width deviation by $45.92\%$ and the variance by $50.34\%$, highlighting its effectiveness in improving corner printing performance.

\begin{table}[h]
\centering
\begin{tabular}{cccc}
\hline
        & Baseline & Optimized printing  & Improvement \\ \hline
Max Error [\SI{}{mm}] & 0.196            & 0.106             & 45.92\%             \\ \hline
Variance [\SI{}{\milli\meter\squared}] & 0.0017            & 0.0008             & 50.34\%             \\ \hline
\end{tabular}
\caption{Quantitative evaluation of corner optimization: comparison of maximum width error and variance between baseline and optimized printing.}
\label{tab:CornerQA}
\end{table}

\subsection{Part printing}
To validate the proposed method, we applied the one-shot optimization framework to print square towers at feed rates of \SI{1600}{\milli\meter\per\minute}, \SI{2400}{\milli\meter\per\minute}, and \SI{3600}{\milli\meter\per\minute}. Baseline towers were sliced conventionally and printed at \SI{900}{\milli\meter\per\minute} and corresponding high speeds for comparative analysis.

\begin{figure}[ht]
     \centering
     \includegraphics[scale=1]{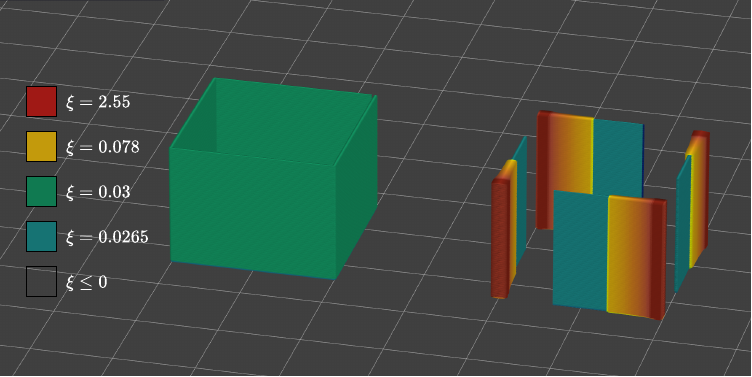}\\
     \caption{Visualization of G-code design for towers at \SI{3600}{\milli\meter\per\minute}: Baseline vs. Optimized Extrusion Commands. The left side shows the baseline tower with constant extrusion, while the right illustrates the optimized tower with dynamic extrusion ratios ($\xi$). Colors represent extrusion ratios along the path, with some sections in the optimized tower exhibiting no extrusion or retraction.}
     \label{fig:TowerGcodeComparsion}
\end{figure}

Figure~\ref{fig:TowerGcodeComparsion} visualizes the baseline and optimized G-code commands for towers printed at \SI{3600}{\milli\meter\per\minute}. The baseline G-code, shown on the left, employs a constant extrusion ratio throughout the print. In contrast, the optimized G-code, depicted on the right, dynamically adjusts the extrusion ratio to synchronize the extrusion and motion systems, thereby enhancing print quality at high speeds.

\begin{figure}[htbp]
  \vspace{-10pt}
  \caption*{Front Views} 
\vspace{-10pt}
  \centering
  \adjustbox{valign=c}{%
    \begin{minipage}{0.27\textwidth}
      \centering
        \begin{subfigure}{\textwidth}
            \centering
            \includegraphics[width=\linewidth]{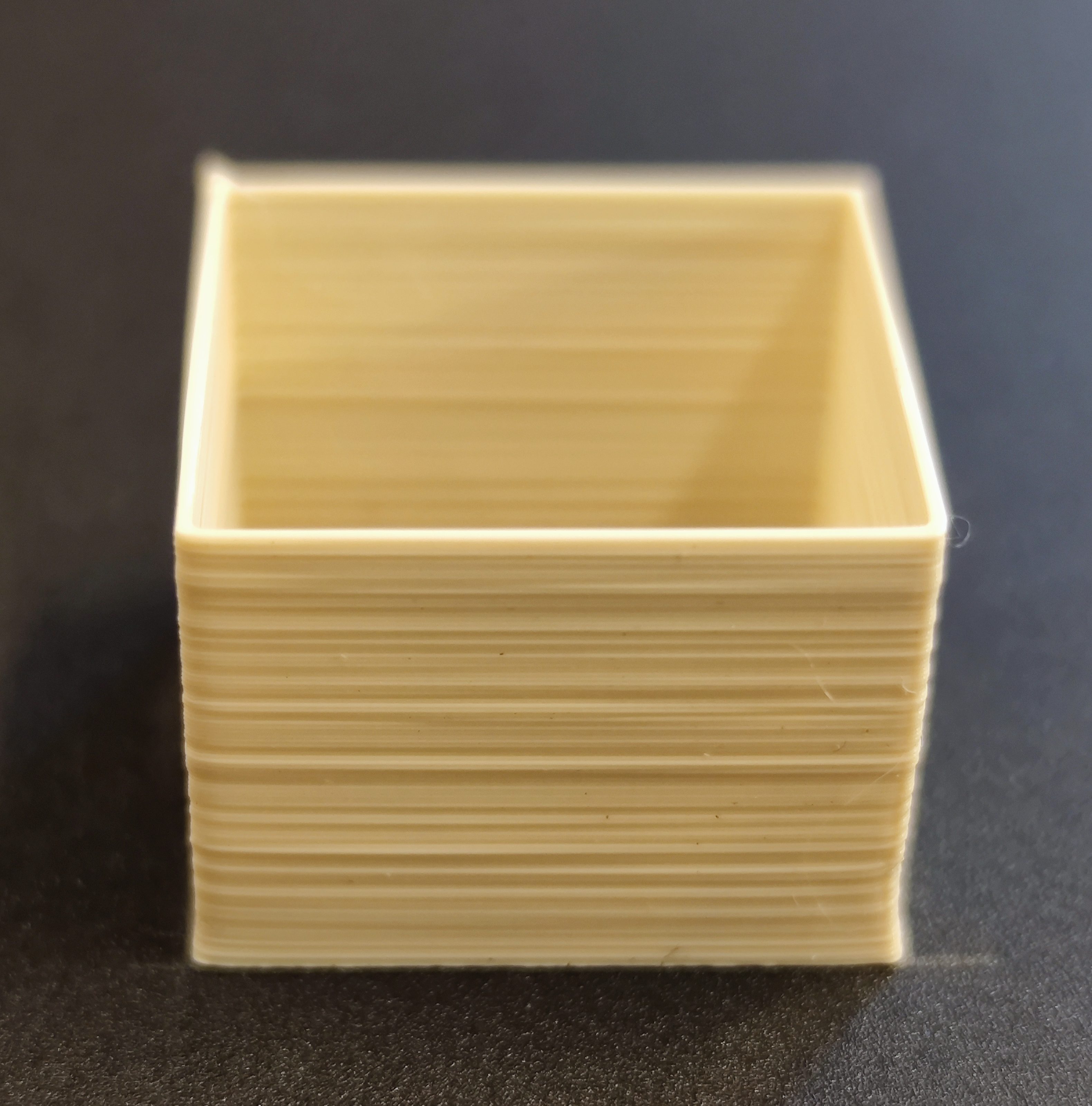} \vspace{-25pt}
            \caption{Baseline tower printed at \SI{900}   
            {\milli\meter\per\minute}}
            \label{fig:900towerprintingfront}
        \end{subfigure}
    \end{minipage}%
  }
  \hfill
  \begin{minipage}[c]{0.27\textwidth}
    \centering
    \begin{subfigure}{\textwidth}
        \centering
        \includegraphics[width=\linewidth]{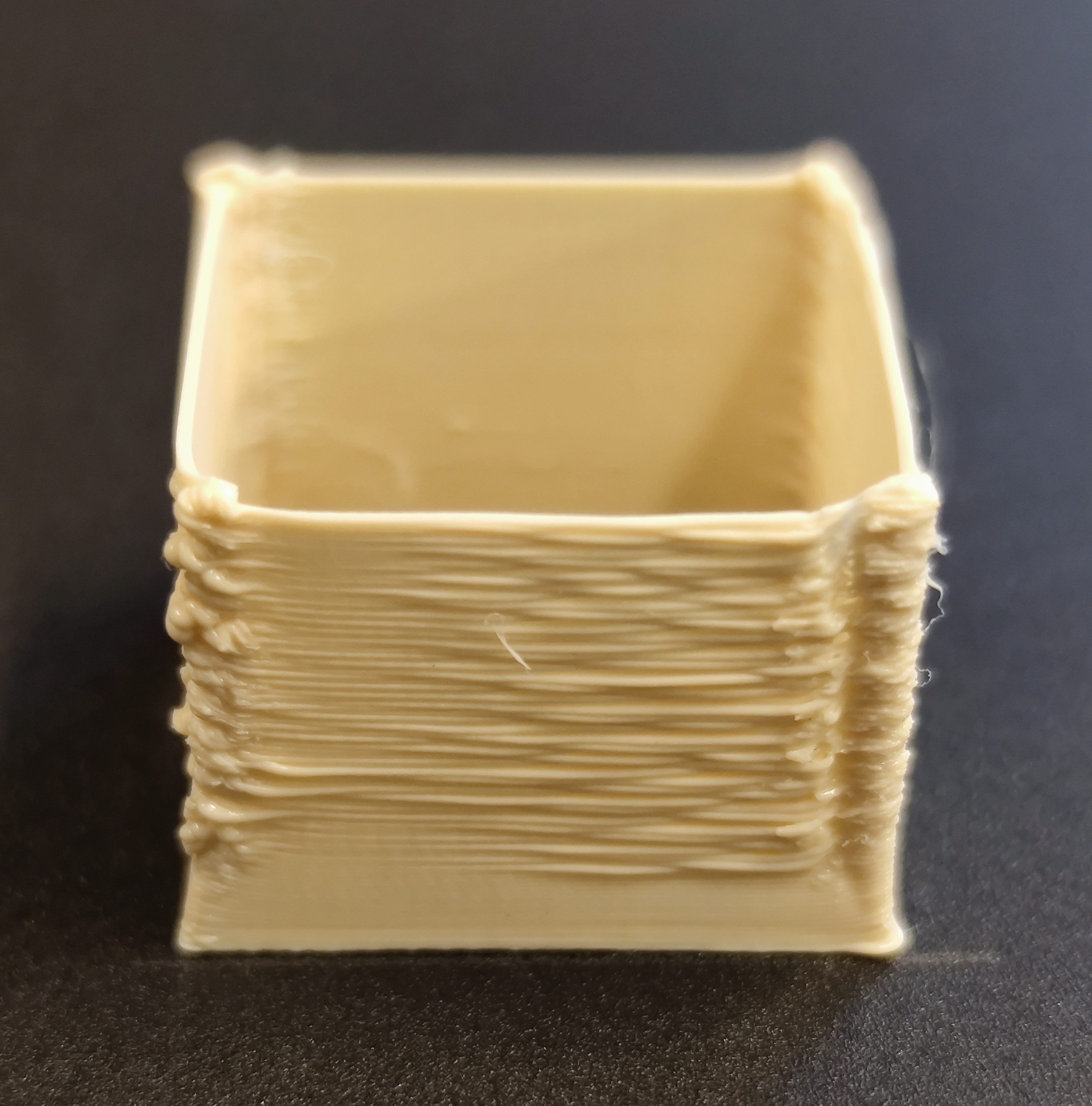}\vspace{-10pt}
        \caption{Baseline tower printed at \SI{3600}{\milli\meter\per\minute}}
        \label{fig:3600towerorginprintingfront}
    \end{subfigure}
    \hfill
    \begin{subfigure}{\textwidth}
        \centering
        \includegraphics[width=\linewidth]{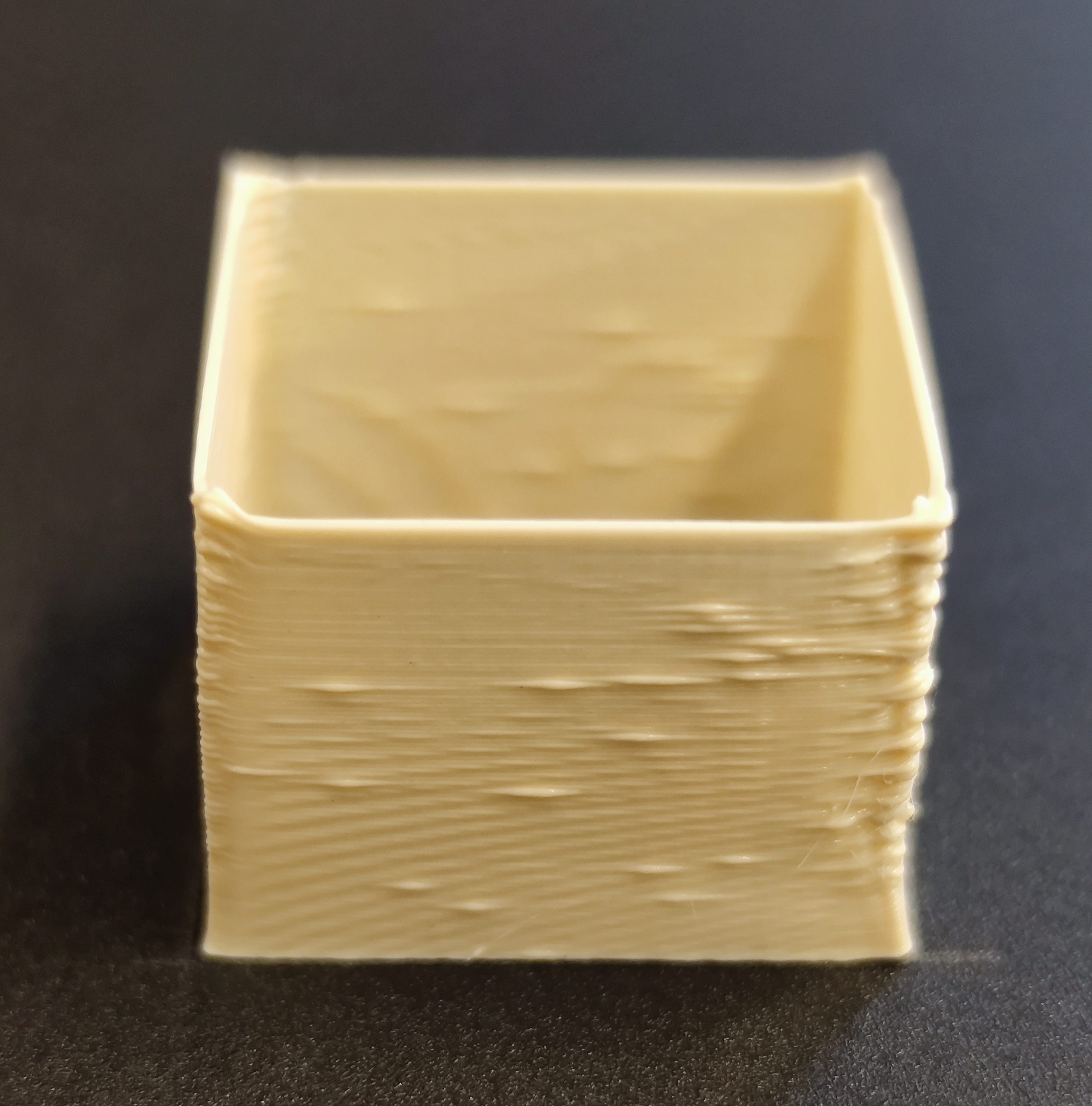}\vspace{-10pt}
        \caption{Optimized tower printed at \SI{3600}{\milli\meter\per\minute}}
        \label{fig:3600toweroptprintingfront}
    \end{subfigure}
  \end{minipage}
  \hfill
  \begin{minipage}[c]{0.27\textwidth}
    \centering
    \begin{subfigure}{\textwidth}
        \centering
        \includegraphics[width=\linewidth]{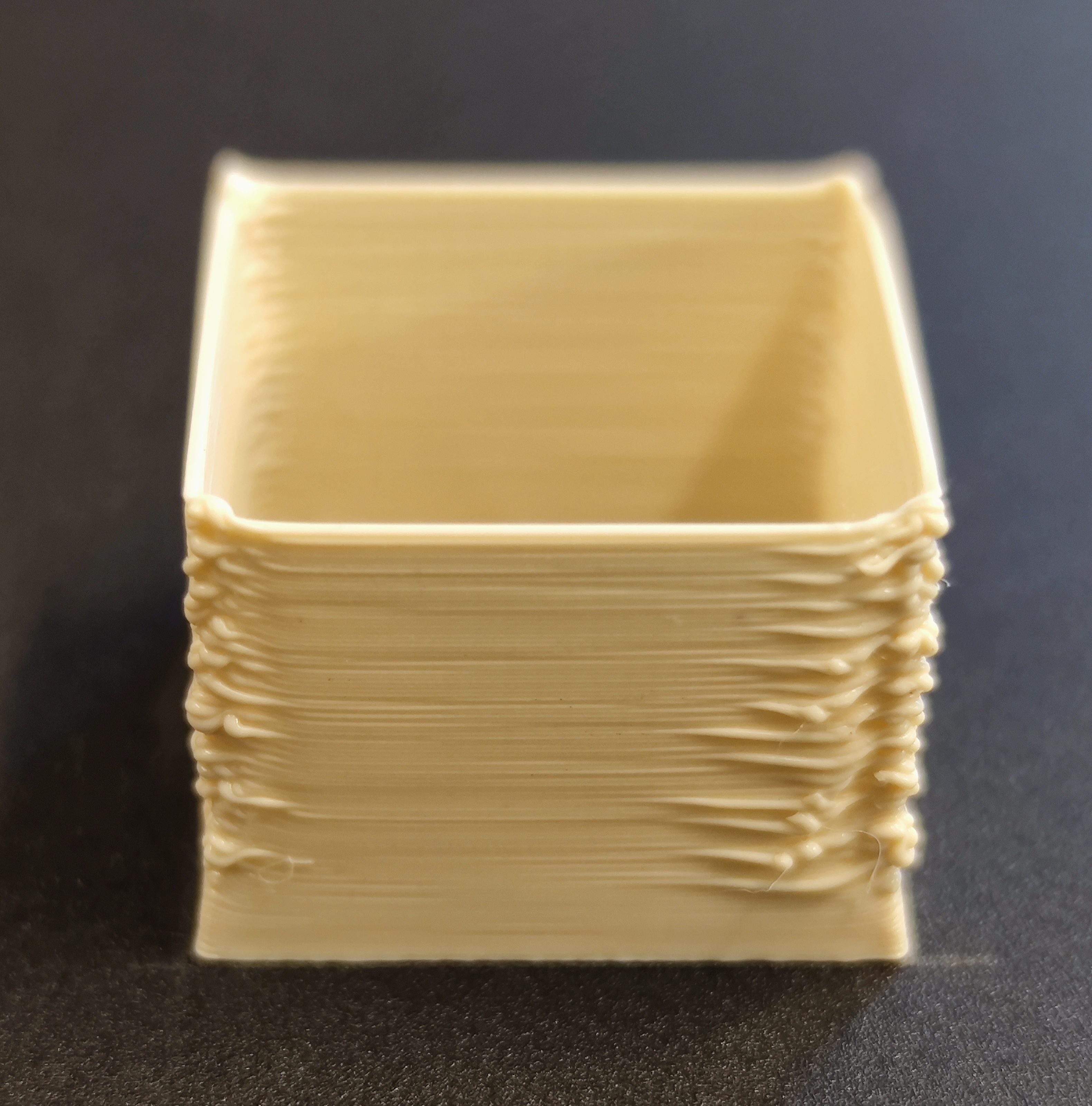}\vspace{-10pt}
        \caption{Baseline tower printed at \SI{2400}{\milli\meter\per\minute}}
        \label{fig:2400towerorginprintingfront}
    \end{subfigure}
    \hfill
    \begin{subfigure}{\textwidth}
        \centering
        \includegraphics[width=\linewidth]{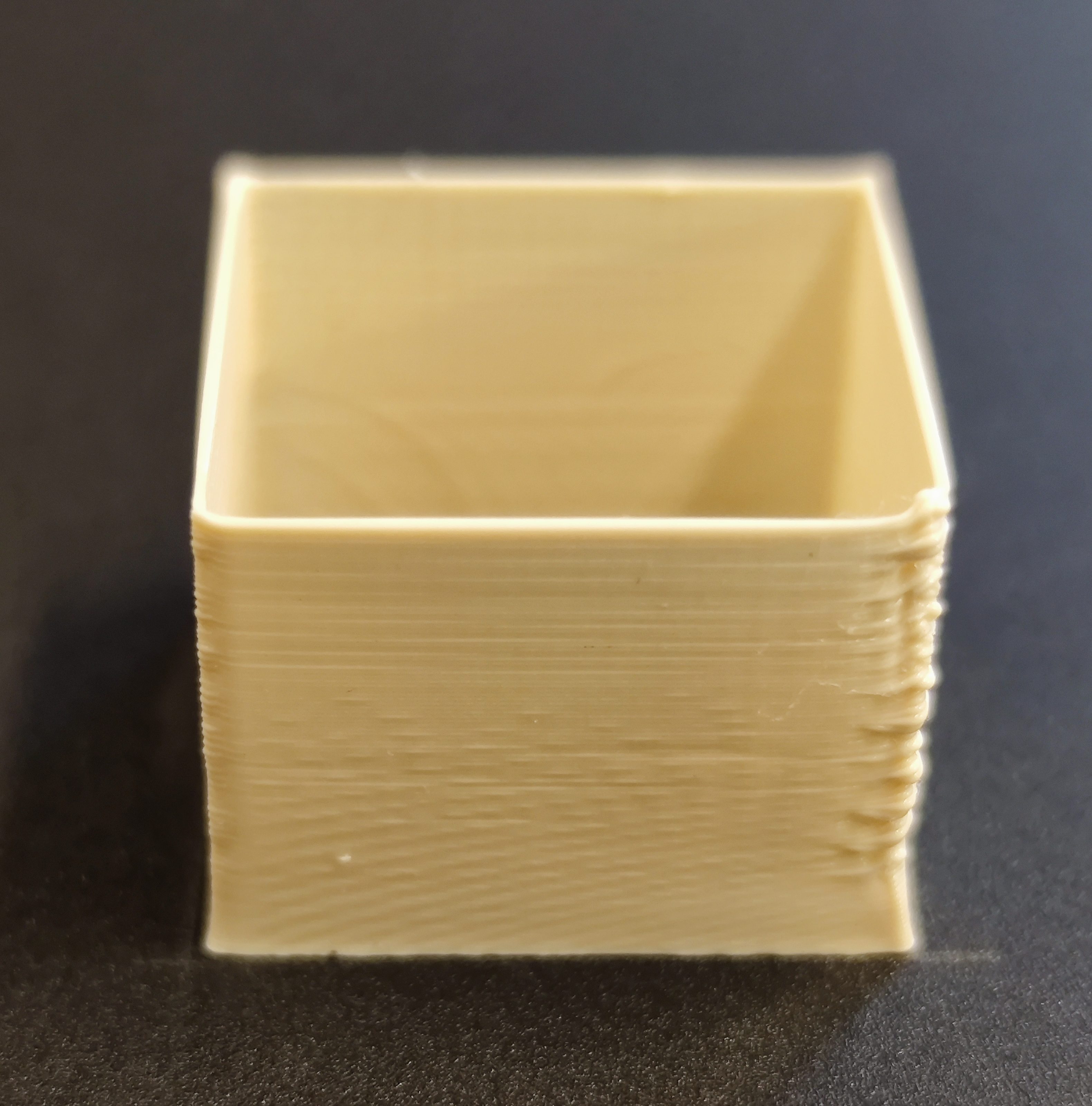}\vspace{-10pt}
        \caption{Optimized tower printed at \SI{2400}{\milli\meter\per\minute}}
        \label{fig:2400toweroptprintingfront}
    \end{subfigure}
  \end{minipage}

  \caption*{Perspective Views} 
  \vspace{-10pt}
  \centering
  \adjustbox{valign=c}{%
    \begin{minipage}{0.27\textwidth}
      \centering
        \begin{subfigure}{\textwidth}
            \centering
            \includegraphics[width=\linewidth]{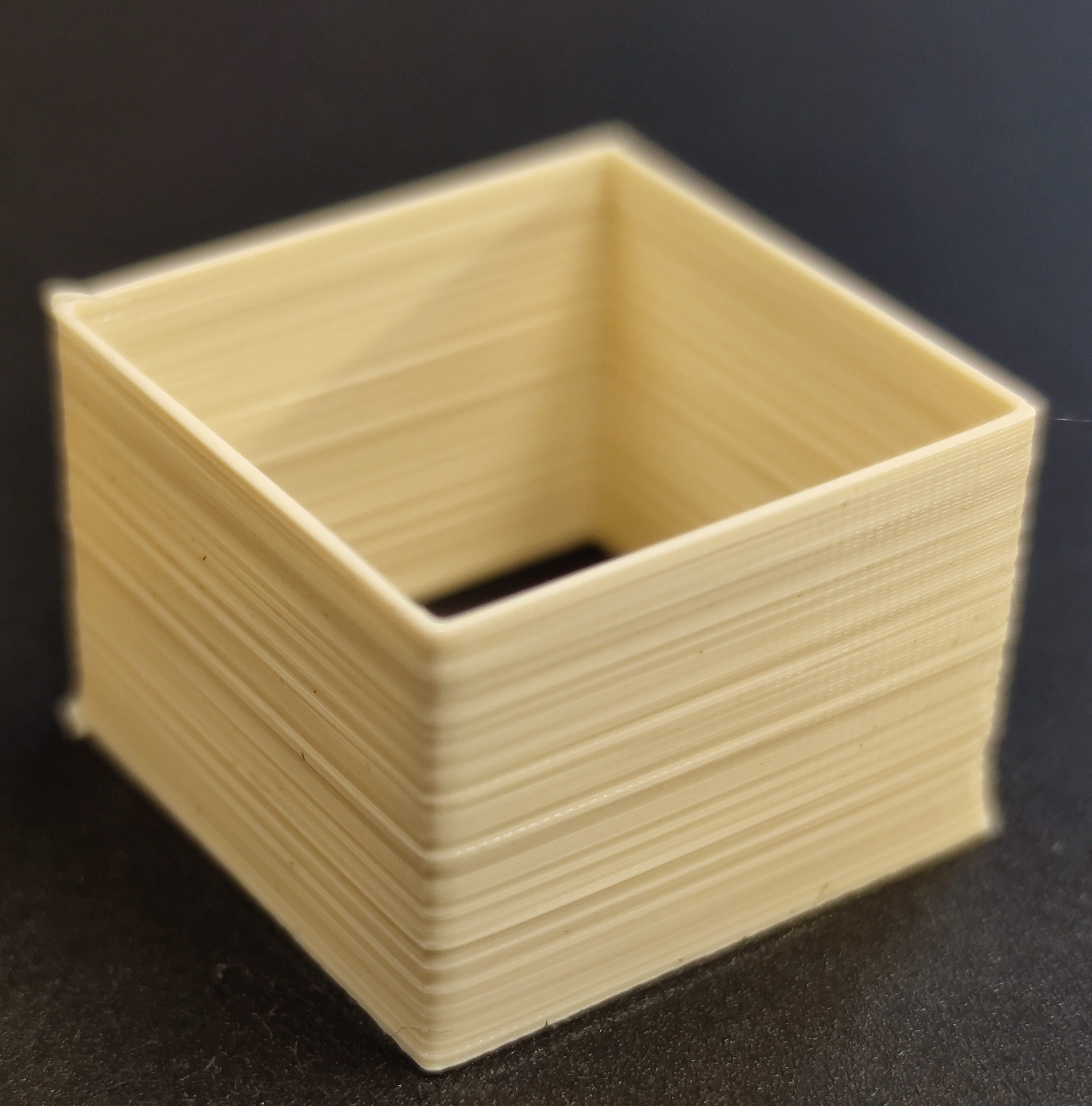} \vspace{-25pt}
            \caption{Baseline tower printed at \SI{900}{\milli\meter\per\minute}}
            \label{fig:900towerprintingside}
        \end{subfigure}
    \end{minipage}%
  }
  \hfill
  \begin{minipage}[c]{0.27\textwidth}
    \centering
    \begin{subfigure}{\textwidth}
        \centering
        \includegraphics[width=\linewidth]{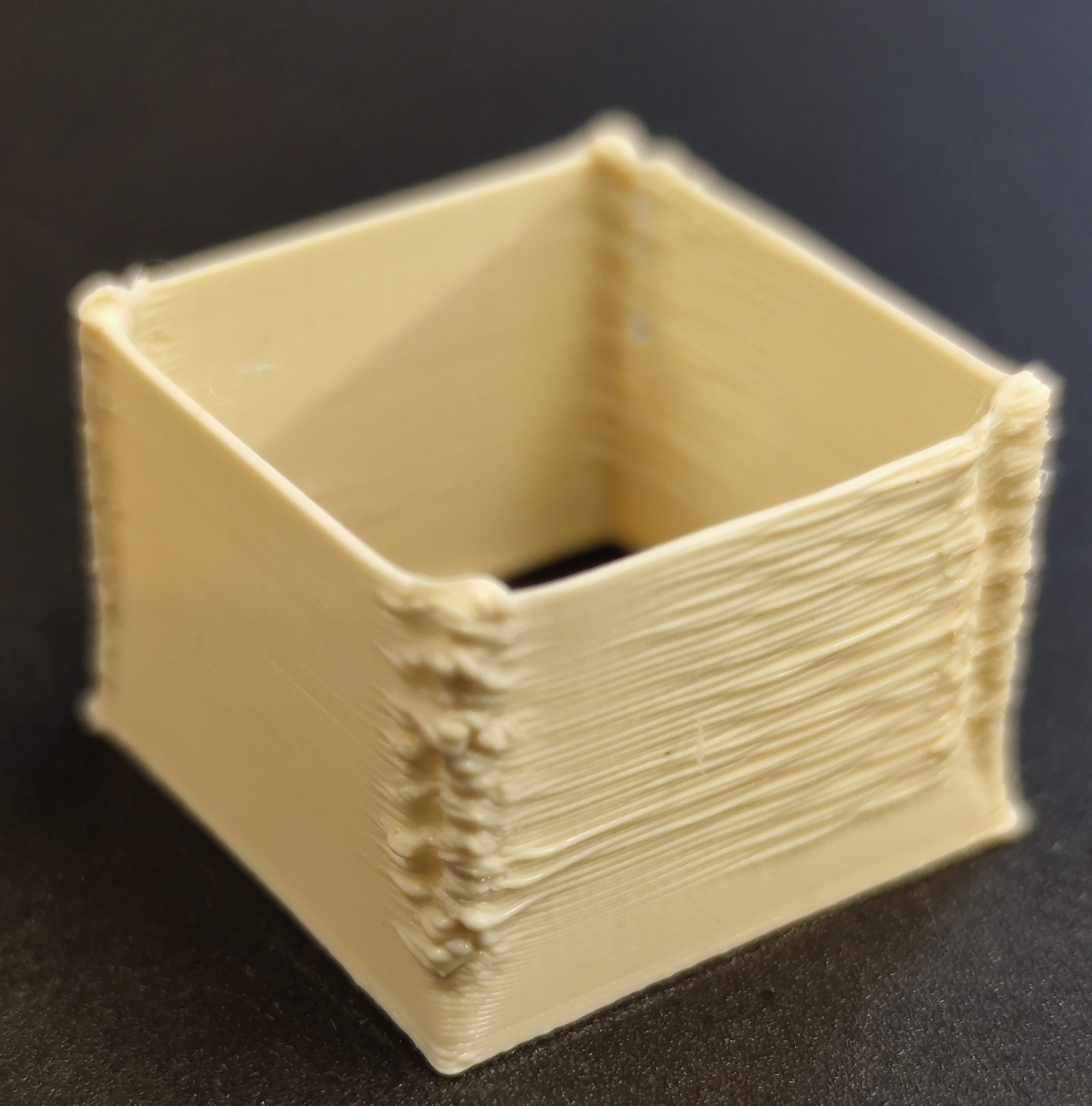}\vspace{-10pt}
        \caption{Baseline tower printed at \SI{3600}{\milli\meter\per\minute}}
        \label{fig:3600towerorginprintingside}
    \end{subfigure}
    \hfill
    \begin{subfigure}{\textwidth}
        \centering
        \includegraphics[width=\linewidth]{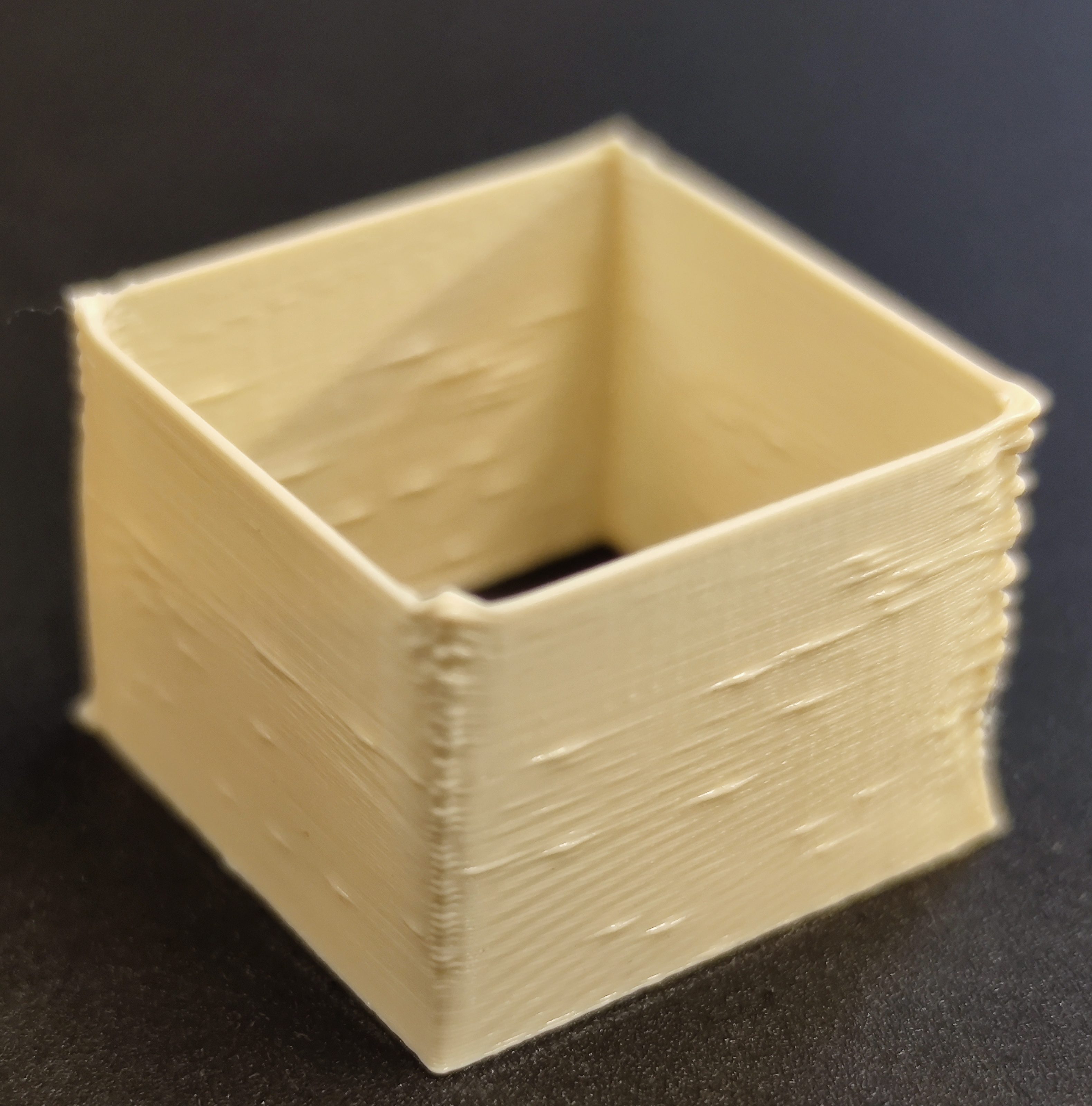}\vspace{-10pt}
        \caption{Optimized tower printed at \SI{3600}{\milli\meter\per\minute}}
        \label{fig:3600toweroptprintingside}
    \end{subfigure}
  \end{minipage}
  \hfill
  \begin{minipage}[c]{0.27\textwidth}
    \centering
    \begin{subfigure}{\textwidth}
        \centering
        \includegraphics[width=\linewidth]{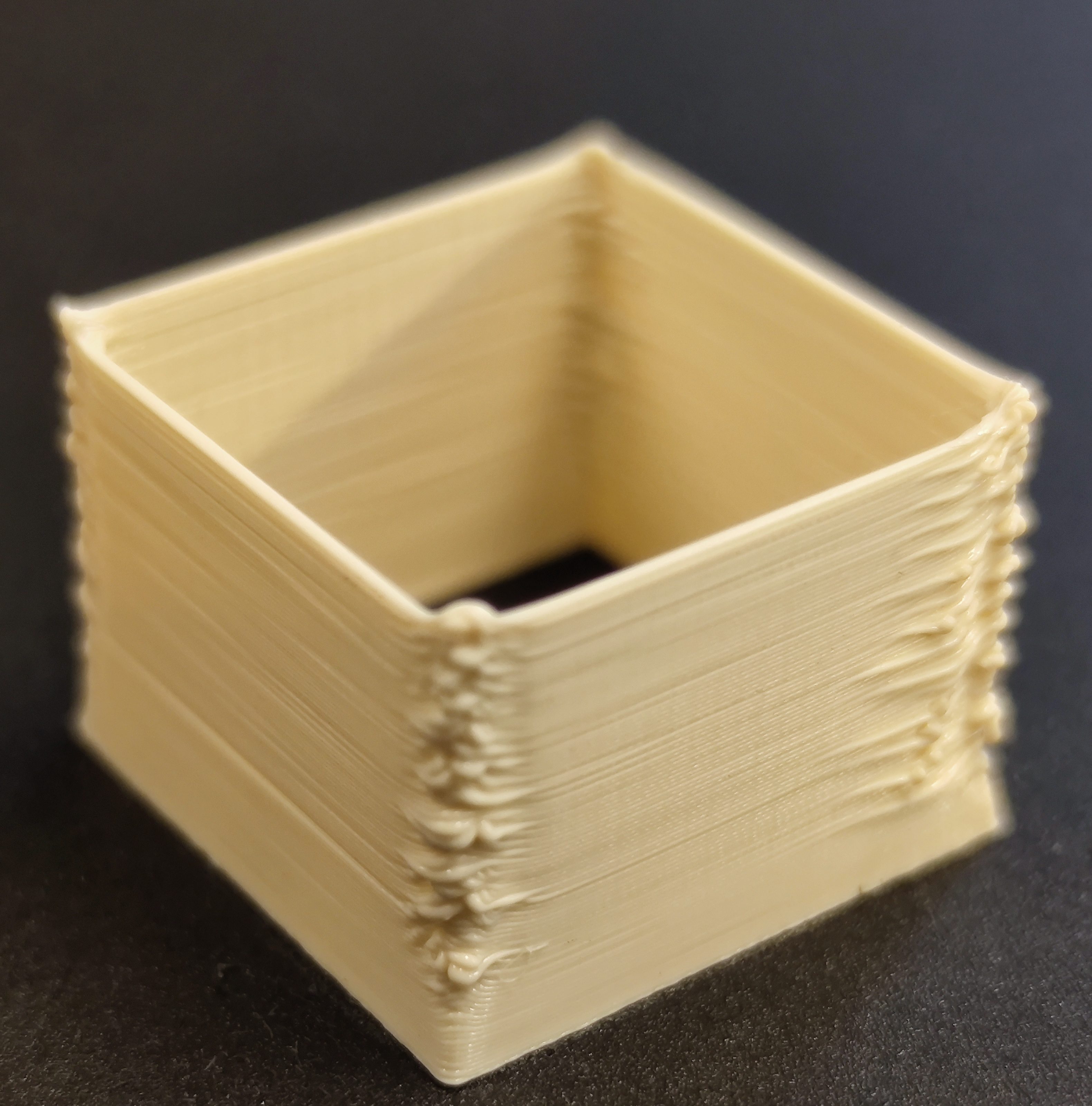}\vspace{-10pt}
        \caption{Baseline tower printed at \SI{2400}{\milli\meter\per\minute}}
        \label{fig:2400towerorginprintingside}
    \end{subfigure}
    \hfill
    \begin{subfigure}{\textwidth}
        \centering
        \includegraphics[width=\linewidth]{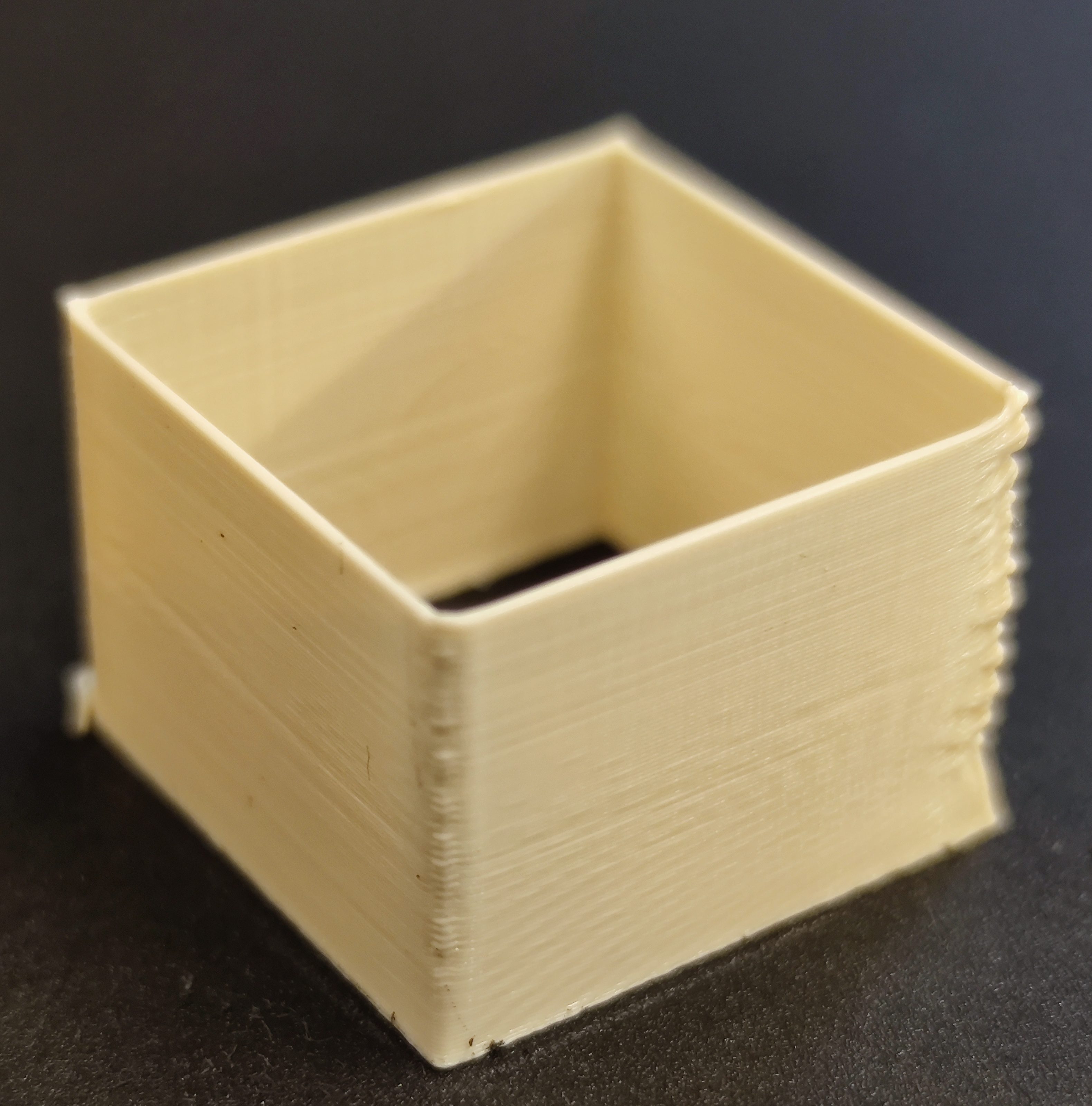}\vspace{-10pt}
        \caption{Optimized tower printed at \SI{2400}{\milli\meter\per\minute}}
        \label{fig:2400toweroptprintingside}
    \end{subfigure}

  \end{minipage}

  \caption{Printed towers comparison}
  \label{fig:TowersComparison}
\end{figure}

The printing results are presented in Fig.~\ref{fig:TowersComparison}, comparing towers printed at different speeds and methods. The baseline and optimized results at \SI{3600}{\milli\meter\per\minute} and \SI{2400}{\milli\meter\per\minute} demonstrate the effectiveness of the proposed method. It can be observed that:

\begin{itemize}
  \item The tower printed at \SI{900}{\milli\meter\per\minute} using the baseline method (Fig.~\ref{fig:900towerprintingfront} and Fig.~\ref{fig:900towerprintingside}) exhibits a smooth surface and no cornering defects, demonstrating that this speed is suitable for high-quality conventional printing.
  \item At higher speeds of \SI{3600}{\milli\meter\per\minute} (Fig.~\ref{fig:3600towerorginprintingfront} and Fig.~\ref{fig:3600towerorginprintingside}) and \SI{2400}{\milli\meter\per\minute} (Fig.~\ref{fig:2400towerorginprintingfront} and Fig.~\ref{fig:2400towerorginprintingside}), the baseline method results in significant print quality deterioration , characterized by rough surfaces and pronounced corner defects.
  \item The proposed optimization method significantly improves print quality at high speeds, yielding smoother surfaces: \SI{3600}{\milli\meter\per\minute}  in Fig.~\ref{fig:3600toweroptprintingfront} and Fig.~\ref{fig:3600toweroptprintingside}, and \SI{2400}{\milli\meter\per\minute}  in Fig.~\ref{fig:2400toweroptprintingfront} and Fig.~\ref{fig:2400toweroptprintingside}, .
  \item While the optimized towers at high speeds do not fully match the quality of the low-speed print, the proposed method effectively balances high print quality with a fourfold reduction in printing time, further validating the effectiveness of the proposed optimization approach.
\end{itemize}

A Micro-Epsilon ILD1220-10 laser distance sensor is employed to scan the tower surfaces and evaluate their roughness, following the validated protocol in \cite{guidetti2023data, balta2021layer}. The side surfaces shown in Fig.~\ref{fig:TowersComparison} were scanned using a raster pattern with a \SI{1}{\milli\meter} pass spacing. The scan data $z_{scan}(x_i,y_i)$ were used to reconstruct the surfaces and fit a plane $\hat{z}(x_i,y_i)$ representing an ideally smooth surface. 

\begin{figure}[h]
    \centering
    \begin{subfigure}{0.49\textwidth}
        \includegraphics[scale=1]{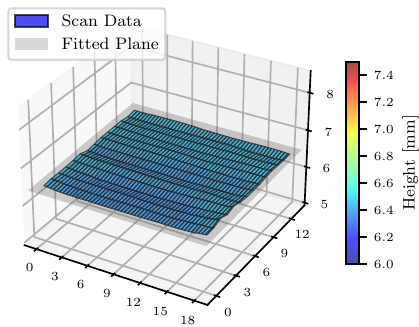}
        \vspace{-5pt} 
        \caption{Scan data and fitted plane}
        \label{fig:F900surface}
    \end{subfigure} 
    \hfill
    \begin{subfigure}{0.49\textwidth}
        \includegraphics[scale=1]{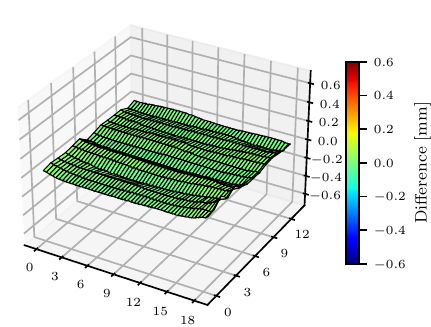}
        \vspace{-5pt} 
        \caption{Deviation from fitted plane}
        \label{fig:F900diff}
    \end{subfigure} \\
    \caption{Surface evaluation of the conventional tower printed at \SI{900}{\milli\meter\per\minute}}
    \label{fig:ScanFigF900}
\end{figure}

\begin{figure}[htbp]
    \centering

    \begin{subfigure}{0.49\textwidth}
        \centering
        \includegraphics[scale=1]{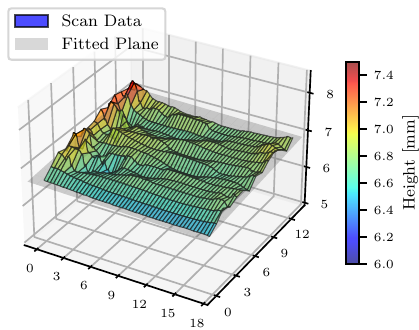}
        \caption{Scan data and fitted plane of the conventional tower}
        \label{fig:Raw3600Sur}
    \end{subfigure} 
    \begin{subfigure}{0.49\textwidth}
        \centering
        \includegraphics[scale=1]{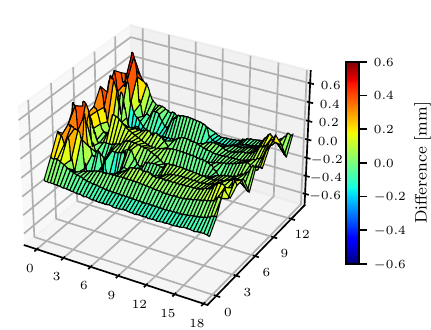}
        \caption{Deviation from fitted plane of the conventional tower}
        \label{fig:Raw3600Diff}
    \end{subfigure} \\

    \begin{subfigure}{0.49\textwidth}
        \centering
        \includegraphics[scale=1]{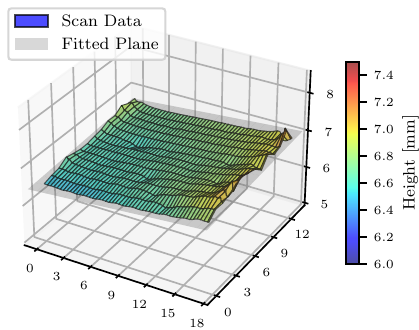}
        \caption{Scan data and fitted plane of the optimized tower}
        \label{fig:Opt3600Sur}
    \end{subfigure} 
    \begin{subfigure}{0.49\textwidth}
        \centering
        \includegraphics[scale=1]{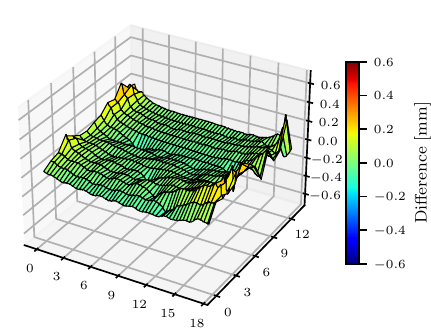}
        \caption{Deviation from fitted plane of the optimized tower}
        \label{fig:Opt3600Diff}
    \end{subfigure} \\

    \caption{Surface evaluation of the conventional and optimized tower printed at \SI{3600}{\milli\meter\per\minute}}
    \label{fig:ScanFig3600}
\end{figure}

\begin{figure}[htbp]
    \centering
    
    \begin{subfigure}{0.49\textwidth}
        \centering
        \includegraphics[scale=1]{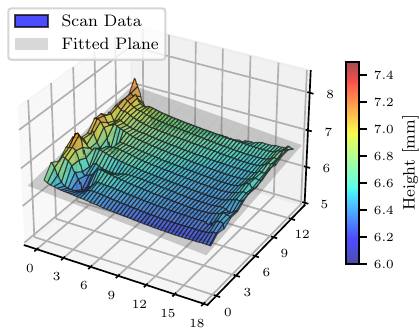}
        \caption{Scan data and fitted plane of the conventional tower}
        \label{fig:Raw2400Sur}
    \end{subfigure} 
    \begin{subfigure}{0.49\textwidth}
        \centering
        \includegraphics[scale=1]{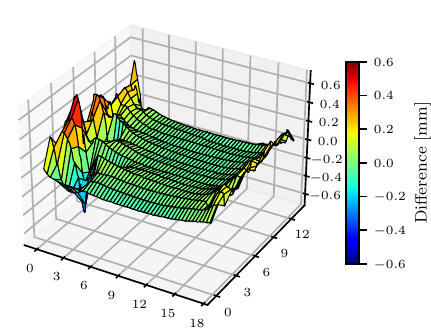}
        \caption{Deviation from fitted plane of the conventional tower}
        \label{fig:Raw2400Diff}
    \end{subfigure} \\

    \begin{subfigure}{0.49\textwidth}
        \centering
        \includegraphics[scale=1]{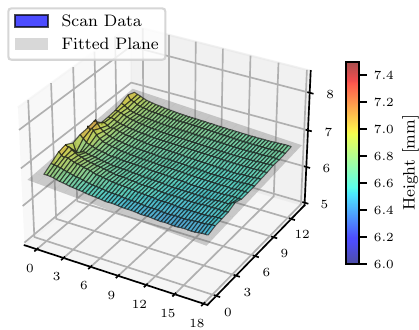}
        \caption{Scan data and fitted plane of the optimized tower}
        \label{fig:Opt2400Sur}
    \end{subfigure} 
    \begin{subfigure}{0.49\textwidth}
        \centering
        \includegraphics[scale=1]{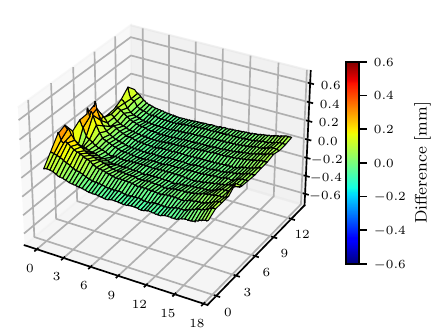}
        \caption{Deviation from fitted plane of the optimized tower}
        \label{fig:Opt2400Diff}
    \end{subfigure} \\
    \caption{Surface evaluation of the conventional and optimized tower printed at \SI{2400}{\milli\meter\per\minute}}
    \label{fig:ScanFig2400}
\end{figure}

Fig.~\ref{fig:ScanFigF900}, Fig.~\ref{fig:ScanFig3600}, and Fig.~\ref{fig:ScanFig2400} present the scan data, fitted surfaces, and deviations for towers printed at \SI{900}{mm\per\minute}, \SI{2400}{mm\per\minute}, and \SI{3600}{mm\per\minute}, respectively. The reconstructions align with the images in Fig.~\ref{fig:TowersComparison}. At \SI{900}{mm\per\minute}, Fig.~\ref{fig:ScanFigF900} demonstrates a smooth surface typical of low-speed printing. In contrast, high-speed baseline printing results in rougher, uneven surfaces, as shown in Fig.~\ref{fig:Raw3600Sur} and Fig.~\ref{fig:Raw2400Sur}, with larger deviations in Fig.~\ref{fig:Raw3600Diff} and Fig.~\ref{fig:Raw2400Diff}. The proposed one-shot optimization significantly improves surface quality, as evidenced by the smoother scan data in Fig.~\ref{fig:Opt3600Sur} and Fig.~\ref{fig:Opt2400Sur}, and reduced deviations in Fig.~\ref{fig:Opt3600Diff} and Fig.~\ref{fig:Opt2400Diff}.

The deviations between the scanned data $z_{scan}(x_i,y_i)$ and the fitted plane $\hat{z}(x_i,y_i)$ were analyzed to quantify surface roughness:
\begin{equation}
  \label{eq:RMSE_printing_part}
   Ra = \sqrt{\frac{1}{n \cdot m} \sum_{i=1}^{n}\sum_{j=1}^{m} (z_{scan}(x_i,y_j) - \hat{z}(x_i,y_j))^2}.
\end{equation}
The calculated roughness values and mean error for towers printed at different speeds and methods are summarized in Table~\ref{tab:surface_roughness_comparison}. 

\begin{table}[htbp]
\centering
\begin{tabular}{ccccc}
\hline
 & \textbf{Metrics} & \textbf{Conventional method} & \textbf{Proposed method} & \textbf{Improvement} \\ \hline
\multirow{2}{*}{\SI{900}{\milli\meter\per\minute}} 
 & $Ra$ [\SI{}{\micro\meter}] & 30 & \multirow{2}{*}{--} & \multirow{2}{*}{--} \\
 & Mean ± std [\SI{}{\micro\meter}] & $30 \pm 20$ & & \\ \hline
 \multirow{2}{*}{\SI{1600}{\milli\meter\per\minute}} 
 & $Ra$ [\SI{}{\micro\meter}] & 60 & 40 & 37\% \\
 & Mean ± std [\SI{}{\micro\meter}] & $50 \pm 60$ & $30 \pm 40$ & 40\% \\ \hline
\multirow{2}{*}{\SI{2400}{\milli\meter\per\minute}} 
 & $Ra$ [\SI{}{\micro\meter}] & 140 & 60 & 51\% \\
 & Mean ± std [\SI{}{\micro\meter}] & $80 \pm 140$ & $40 \pm 60$ & 55\% \\ \hline
\multirow{2}{*}{\SI{3600}{\milli\meter\per\minute}} 
 & $Ra$ [\SI{}{\micro\meter}] & 130 & 70 & 43\% \\
 & Mean ± std [\SI{}{\micro\meter}] & $90 \pm 130$ & $50 \pm 70$ & 48\% \\ \hline
\end{tabular}
\caption{Comparison of surface by roughness ($Ra$) and mean ± standard deviation between conventional and proposed methods at different printing speeds. Values are reported in micrometers (\SI{}{\micro\meter}) rounded to a resolution of 10\SI{}{\micro\meter}.}
\label{tab:surface_roughness_comparison}
\end{table}

The one-shot optimization framework demonstrates significant improvements in surface quality for high-speed printing. The surface roughness is reduced at least by $37\%$, compared to the conventional approach. Notably, the optimized towers printed at \SI{2400}{\milli\meter\per\minute} and \SI{3600}{\milli\meter\per\minute} achieve surface quality comparable to that of unoptimized towers printed at \SI{1600}{\milli\meter\per\minute}. Furthermore, applying the optimization to \SI{1600}{\milli\meter\per\minute} printing results in surface quality similar to that of unoptimized printing at \SI{900}{\milli\meter\per\minute}. These results highlight the effectiveness of the proposed optimization in bridging the quality gap between high-speed and low-speed printing, enabling faster production without compromising print quality.

\section{Conclusions and future work} \label{sec:conclusion}

This paper has introduced a practical end-to-end method to optimize G-code to enhance the performance of conventional 3D printers operating at high speeds. 
A modeling approach for extrusion and corner printing dynamics has been proposed and utilized to optimize and synchronize the motion of the printhead with the filament extrusion system, ensuring enhanced printing performance. The proposed one-shot method only requires phone cameras to achieve the identification of extrusion and cornering dynamics without hardware modifications, and a optimized G-code can be generated to improve print quality at high speeds for a wide range of 3D printers.

The proposed method leverages a single phone-camera photograph to achieve accurate one-shot system identification, with an RMSE of \SI{0.03}{\milli\meter} compared to ground truth. By integrating extrusion and cornering dynamics into an optimal constrained control framework, the method significantly enhances high-speed printing performance. It reduces line width tracking errors, mitigates high-speed corner defects, and enhances surface quality by reducing roughness. These results highlight the practicality and effectiveness of the approach, requiring only two simple calibration patterns and minimal user effort.

This work demonstrates the potential of low-cost and user-friendly solutions based on ubiquitous phone cameras to improve the 3D print quality for a wide range of 3D printers users. In future works, we plan to apply phone-camera-based measurements to the optimization of more printing cases and scenarios (e.g., leveling, retraction, etc), and to develop more robust image process methods. We envision to package these solution in a phone application that can benefit a wide audience of 3D printing users and enthusiasts.  

\section*{Funding}

This work was supported by the Swiss Innovation Agency (Innosuisse, grant \textnumero 102.617) and by the Swiss National Science Foundation under NCCR Automation (grant \textnumero 180545).

\section*{Acknowledgments}

We acknowledge the support of NematX AG for conducting this work.



 \bibliographystyle{elsarticle-num} 
 \bibliography{cas-refs}





\end{document}